\documentclass[12pt,preprint]{aastex}
\usepackage{graphicx}
\usepackage{amssymb}
\usepackage{color}
\usepackage{natbib}
\usepackage{amsmath}
\newcommand{\del}{\nabla}
\newcommand{\divflux}{\bf{\del}\cdot {\rm F}}

\newcounter{saveeqn}
\newcommand{\alphaeqn}{\setcounter{saveeqn}{\value{equation}}%
  \stepcounter{saveeqn}\setcounter{equation}{0}%
  \renewcommand{\theequation}
      {\mbox{\arabic{saveeqn}\alph{equation}}}}
\newcommand{\reseteqn}{\setcounter{equation}{\value{saveeqn}}%
  \renewcommand{\theequation}{\arabic{equation}}}

\shorttitle{Radiative Cooling with Realistic Opacities}
\shortauthors{Boley et al.}

\begin{document}
\title{The Thermal Regulation of Gravitational Instabilities in Protoplanetary Disks III. Simulations with Radiative Cooling and Realistic Opacities}

\author{A.\ C.\ Boley} 
\affil{Indiana University, Astronomy Department, 727 E.\ Third St., Bloomington, IN 47405-7105}
\email{acboley@astro.indiana.edu}

\author{ Annie C.\ Mej\'ia}
\affil{Department of Astronomy, University of Washington, Box 351580, Seattle, WA 98195-1580}

\author{Richard~H.~Durisen, Kai Cai}
\affil{Indiana University, Astronomy Department, 727 E.\ Third St., Bloomington, IN 47405-7105}

\author{Megan~K.~Pickett}
\affil{Department of Chemistry and Physics, Purdue University Calumet, 2200 169th St., Hammond, IN 46323-2094}
\author{Paola D'Alessio}
\affil{Instituto de Astronom'a, UNAM, Apartado Postal 70-264, Ciudad Universitaria, 04510 M\'exico DF, M\'exico}

\begin{abstract}
This paper presents a fully three-dimensional radiative hydrodymanics simulation with realistic opacities for a gravitationally unstable $0.07~\rm M_{\odot}$ disk around a 0.5 M$_{\odot}$ star.  We address the following aspects of disk evolution: the strength of gravitational instabilities under realistic cooling, mass transport in the disk that arises from GIs, comparisons between the gravitational and Reynolds stresses measured in the disk and those expected in an $\alpha$-disk, and comparisons between the SED derived for the disk and SEDs derived from observationally determined parameters.   The mass transport in this disk is dominated by global modes, and the cooling times are too long to permit fragmentation for all radii.  Moreover, our results suggest a plausible explanation for the FU Ori outburst phenomenon.

\end{abstract}


\section{INTRODUCTION\label{intro}}

Accurately simulating cooling in protoplanetary disks is fundamental to modeling disk evolution.  If a disk with enough mass becomes sufficiently cold, gravitational instabilities (GIs) set in and produce nonaxisymmetric structure, which provides mass and angular momentum transport through long-range torques. The strength of these GIs and their efficiency in transporting mass depend on the cooling rate \citep{lodato_rice2004,mejia2005}.  Moreover, if the cooling rates are very high, the stress in the disk can become large enough to cause fragmentation \citep{gammie2001,mnras364l56} and possibly induce formation of bound protoplanetary clumps \citep{boss1997,boss1998,boss2001,boss2002}.   Because cooling rates can be a principal determinant of disk evolution \citep{pickett1998,pickett2000}, we here investigate the evolution of a 0.07 M$_{\odot}$ disk surrounding a 0.5 M$_{\odot}$ star by using three-dimensional radiative hydrodynamics with realistic opacities.

A gas disk is both massive and cold enough for GIs to set in when the \citet{toomreQ} parameter
\begin{equation}Q=\frac{c_s\kappa_e}{\pi G \Sigma}\label{eq1}\end{equation}
becomes less than about 1.5 \citep{durisen_r2003}; here, $c_s$ is the sound speed, $\kappa_e$ is the epicyclic frequency, and $\Sigma$ is the gas surface density.  The strength of these GIs strongly depends on the thermal physics of the gas \citep[see][]{pickett2000} and on disk cooling.  To date, simulations of gravitationally unstable disks exhibit a variety of evolutionary behaviors. Many of the differences can be linked to the equation of state and the techniques used to model radiative cooling \citep[for a review, see][]{durisen_ppv_chapter}.  For example, a global cooling time $t_{\rm cool}=$ constant everywhere is used in the simulations of \citet[hereafter Paper I]{pickett1998,pickett2000,pickett2003} and \citet[hereafter Paper II]{mejia2005}.  Their simulations show that mass and angular momentum transport are dominated by low-order modes. When their disks are in the {\it asympototic phase} (see Paper II), where shock heating from gravitational instabilities balances cooling, a simple $\alpha$-disk description \citep{alphadisk} inaccurately describes the disk evolution.  However, if a cooling time that is dependent on the the gas orbital speed is enforced, i.e., $t_{\mathrm{cool}}\Omega=\rm constant$  \citep[e.g.,][]{gammie2001,lodato_rice2004,mayer2004_aspc}, mass and angular momentum transport can be described by an $\alpha$-disk model with an
\begin{equation}\alpha=\left( \left| \frac{\mathrm{d}\ln \Omega}{\mathrm{d}\ln r} \right|^2 \gamma '
\left(\gamma '  - 1\right) t_{\mathrm{cool}} \Omega \right)^{-1},\label{eq2}\end{equation}
as discussed in \citet{gammie2001}.  Here, $\gamma '$ is the two-dimensional ratio of specific heats and equals $3-2/\gamma$ in the strongly self-gravitating limit and $\left(3\gamma-1\right)/\left(\gamma+1\right)$ in the non-self-gravitating limit, where $\gamma$ is the three-dimensional ratio of specific heats. A $t_{\mathrm{cool}}\Omega=\rm constant$ leads to an $\alpha$-disk-like behavior in the asymptotic phase because the stress in the disk that results from GIs must lead to heating that balances $t_{\rm cool} \Omega = \rm constant$ cooling. Similarly, prescribing a global cooling time should lead to a dominance in disk heating by lower-order modes, which provide significant long-range torques over most of the disk.  Let $ P_{\rm rot} = {\rm  rotation~period} = 2\pi/\Omega$. For either prescription, when $t_{\mathrm{cool}}$ becomes less than about 3/$\Omega\approx P_{\rm rot}/2$, for a $\gamma=5/3$ gas, the disk fragments \citep{gammie2001,rice2003,mejia2005}; simulations in \citet{mnras364l56} suggest that this critical cooling time for fragmentation may be closer to 6/$\Omega\approx P_{\rm rot}$ for $\gamma=5/3$ and that it increases to about 12/$\Omega\approx 2 P_{\rm rot}$ for $\gamma=7/5$. Because different cooling prescriptions lead to different GI behavior and disk evolution, radiative hydrodynamics with realistic opacities, a realistic equation of state (EOS), and proper boundary conditions must be used to address how real disks will behave.

Some work has already been done on the protoplanetary disk problem with radiative hydrodynamics.  \citet{nelson_benz_ruzmaikina2000} used a radiative cooling scheme for their two-dimensional hydrodynamics simulations by extrapolating vertical structure from their thin disks by assuming polytropic vertical hydrostatic equilibrium.  They also created SEDs from their data and compared their disks with SEDs derived from observed parameters.  \citet{boss2001} implemented fully 3-D radiative diffusion \citep{bodenheimer1990}. \citet{mejiaphd} employed fully 3-D flux-limited diffusion with different boundary conditions.  The \citet{boss2001} and the \citet{mejiaphd} results are quite different.  \citet{boss2001,boss2002,boss2004,boss2005} argues that radiative physics permits disk fragmentation because convection makes the cooling times very short.  This fragmentation leads to direct planet formation by disk instability.  Furthermore, \citet{boss2002} argues that the disk cooling rate is insensitive to metallicity, and that the observed metallicity-planet correlation \citep{fischer_valenti2005}, which is often considered support for core-accretion plus gas capture \citep{pollack1996} when taken at face value, can be explained by metallicity's effect on inward planet migration rates \citep{boss2005}.  On the other hand, the simulation presented here along with the results of \citet{cai_letter_2006}, which both use the radiative scheme developed by \citet{mejiaphd}, show that cooling times are too long to permit direct planet formation by disk fragmentation.  However, GIs may still aid planet formation.  \citet{durisen2005}, following work by \citet{haghighipour_boss2003},  suggest that rings formed by GIs could lead to accelerated core-accretion plus gas capture. In fact, \citet{rice2004} find that solids are concentrated into the spiral arms of GI active disks.  

The subject of Paper II was a series of simulations with a constant global cooling time (CCT) prescription.  This paper presents a disk simulation evolved under realistic radiative cooling with the \citet{mejiaphd} scheme, and it is an extension with new analyses of the work presented in \citet{mejiaphd}. Section 2 describes the radiative cooling (RC) algorithm used in the hydrodynamics code, while \S 3 presents the simulation setup, evolution, and results.  Comparisons between the radiative cooling and constant cooling (Paper II) simulations are made in \S 4.  Section 5 discusses the relevance of our work to real disks and compares our results to work by other groups. We  summarize our conclusions in \S 6.

\begin{figure}[ht!]
\begin{center}
\includegraphics[width=12cm]{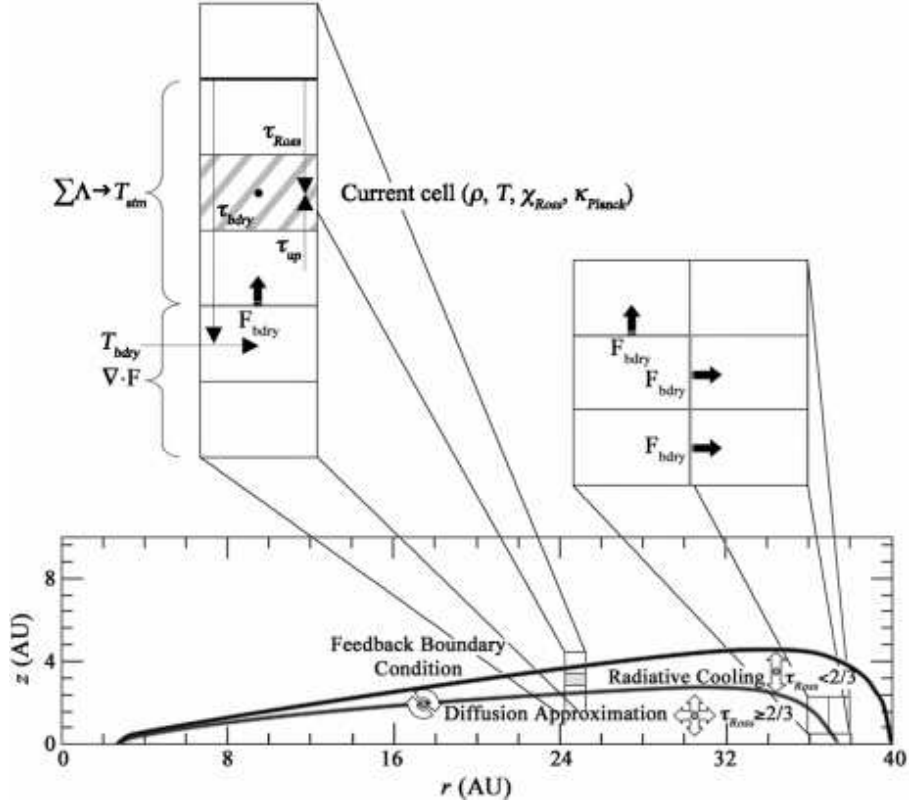}
\caption{Scheme showing the two optical regions of the initial disk (to scale).   The left inset represents part of a column where both the atmosphere and the interior of the disk are shown.  The rectangle with the diagonal pattern is the current atmospheric cell at which the cooling is being calculated using equation (3).  Optical depths, temperatures, and the boundary fluxes are represented by arrows.  The right inset shows the boundary conditions for the last interior column in the radial direction, where the radial boundary fluxes are set equal to the vertical boundary flux.}
\label{f1}
\end{center}
\end{figure}

\section{HYDRODYNAMICS}
The three-dimensional hydrodynamics code with self-gravity is the same as that used in Papers I and II and is described in detail in \citet{pickett1995}, \citet{pickett1998,pickett2000}, and \citet{mejiaphd}. The hydrodynamics code solves Poisson's equation, an ideal gas equation of state, and the equations of hydrodynamics \citep{yang1992} in conservative form on a uniform cylindrical grid ($r$, $\phi$, $z$).  The code computes the source and flux terms \citep{norman_winkler1986} separately in an explicit, second-order time integration \citep{vanalbada_vanleer_roberts1982,christodoulouphd,yang1992}, where the advective terms are calculated with a van Leer scheme  \citep{norman_winkler1986}.  The energy equation has the form \citep{williamsphd,pickett2000,mejiaphd}
\begin{equation} \frac{\partial \epsilon^{1/\gamma} }{\partial t} 
+ {\bf \nabla}\cdot \left( \epsilon^{1/\gamma} {\bf v} \right) = \frac{1}{\gamma} \epsilon^{1/\gamma-1} \left( \Gamma -
\Lambda- \divflux \right),\label{eq3} 
\end{equation}
where $\epsilon$ is the internal energy density and the heating term $\Gamma$ includes the effects of shock heating by artificial bulk viscosity through a second-order Neumann \& Richtmeyer scheme \citep[see][]{pickett1995}.   This artificial viscosity ensures that the jump conditions are satisfied by adding the correct amount of entropy to the gas. For more details on the implemented AV scheme, we refer the reader to \citet{pickett1995} and to \citet{pickett2000}. The cooling terms $\Lambda$ and $\divflux$, which are described below, represent the local radiative cooling in the atmosphere of the disk and in the interior of the disk, respectively.

\subsection{Radiative Cooling}

Our radiative physics scheme employs two approximations for the divergence of the flux
\begin{equation} {\divflux} = \int_{4\pi}  \rho\kappa
\left(S - J \right)~\mathrm{d}\Omega,\label{eq4}\end{equation}
where $\kappa$ is the mass absorption coefficient, $\rho$ is the density, $S$ is the source function, and $J$ is the mean intensity.  The two limits are fit together with an Eddington-like boundary condition.  When the optical depth $\tau$ is very large, equation (4) takes a diffusion form, which is approximated in finite difference form as
\begin{equation} {\divflux} = \frac{1}{r} \frac{\Delta\left(r\mathrm{F}_r\right)}
{\Delta r} + \frac{1}{r}\frac{\Delta \mathrm{F}_{\phi} }{\Delta \phi} + \frac{\Delta \mathrm{F}_z}
{\Delta z}.\label{eq5}\end{equation}
In this limit, the radiative flux is calculated by flux-limited diffusion, as described below, and Rosseland mean opacities, where $\kappa=\chi_{\rm Ross}$ (see Appendix A), are used.  When the mean intensity becomes negligible and a parcel of gas is allowed to radiate as much as its emissivity allows, equation (4) takes the form
\begin{equation} {\divflux} =4\rho\kappa_{\rm Planck}\sigma T^4,
\label{eq6}\end{equation}
where the Planck mean opacity $\kappa_{\rm Planck}$ is appropriate in this limit and the emissivity $B(T)\kappa_{\rm Planck} = \left(\sigma T^4/\pi\right)\kappa_{\rm Planck}$.

We define two regions of the disk, the interior and the atmosphere, through the Rosseland optical depth, which is integrated vertically downward toward the disk midplane.  The interior is defined for cells with $\tau_{\rm Ross} \ge 2/3$, and in this region the cooling algorithm uses flux-limited diffusion.  When $\tau_{\rm Ross} < 2/3$, which defines the atmosphere cells, a different approach is required.  A real atmosphere would not cool according to equation (6), but it would absorb photons emerging from the interior of the disk.  Therefore, we couple the atmosphere to the interior by defining the {\it net} atmosphere cooling as
\begin{equation} \Lambda=4\rho\kappa_{\rm Planck}\sigma T^4 -\rho \chi_{\rm Ross}\mathrm{F_{bdry}}
e^{ -\tau_{\rm up}},\label{eq7}
 \end{equation}
where the first term on the right-hand side indicates pure radiative cooling for an atmospheric cell given by its own temperature and the second term on the right-hand side accounts for heat gained by the atmospheric cell from the underlying disk.  The second term is nonzero if there is an upward energy flux $\rm F_{bdry}$ at the boundary between the atmosphere and the interior of the disk.  The boundary flux is diminished exponentially by the Rosseland optical depth, which is measured from the upper boundary cell (labeled $\tau_{\rm up}$ in Fig.\ 1).  If the boundary flux is negative (downward) due to shocks and/or irradiation in the atmosphere, the second term on the right in equation (7) is made zero such that the cell can cool as much as its emissivity allows. Without the second term on the right-hand side of (7), our atmosphere regions contract into a single cell within a few outer rotations.

The boundary flux provides the boundary condition between the atmosphere and the interior of the disk by defining the balance between the energy leaving the disk interior through the atmosphere and the energy gained from the atmosphere.  If no atmospheric heating is assumed and all the energy that flows through the boundary is radiated with an Eddington effective temperature fitted to some boundary temperature $T_{\rm bdry}$ and the corresponding Rosseland optical depth at $\tau_{\rm bdry}$, both of which are measured at the center of the first disk interior cell in a column, the flux at the boundary would be
\begin{equation}{\rm F_{\rm bdry}}=\frac{ 4 \sigma T_{\rm bdry}^4 }{3\left( \tau_{\rm bdry} + 2/3 \right) },
\label{eq8}\end{equation}
similar to the equation for Eddington gray atmosphere.   To add the contribution from the atmosphere, it is assumed that half the radiation emitted by the atmosphere leaves the top of the atmosphere and half shines on the disk.  Then the boundary flux becomes
\begin{equation}{\rm F_{\rm bdry}}=\frac{ 4 \sigma \left( T_{\rm bdry}^4-T_{\rm atm}^4\right)
 }{3\left( \tau_{\rm bdry} + 2/3 \right) },\label{eq9}\end{equation}
where the atmospheric $T_{\rm atm}$ is a measure of the downward flux into the interior of the disk from the atmosphere, and it is given by half the sum of the atmospheric cooling
\begin{equation} T^4_{\rm atm}
=2\sum_{\rm atm}\rho\kappa_{\rm Planck} T^4 \Delta z. \label{eq10}\end{equation}
Equation (10) only accounts for the first term on the right in equation (7).  In the 3-D code, $T_{\rm atm}$ is calculated first, then F$_{\rm bdry}$, and finally $\Lambda$.  Because the atmospheric cells can be heated by shocks generated by GI-activity \citep{boleyshockbores} and because these routines are also used to treat outside irradiation in other calculations \citep{cai_letter_2006}, it is possible for $T_{\rm atm}^4 > T_{\rm bdry}^4$, which makes $\rm F_{bdry}<0$.  These somewhat cumbersome procedures are necessary to fit the explicitly modeled upper, low-$\tau_{\rm Ross}$ region to the inner, high-$\tau_{\rm Ross}$ region through an Eddington-like solution across the bounding cell, where the Eddington-like solution fitted across this cell has to be modified, as in equation (9), to include downward radiation from above the cell.
 
The discussion above only applies to the vertical boundary condition in grid columns that contain both atmosphere and interior cells, like the first inset in Figure 1.  However, another boundary condition must be set to the fluxes between cells when one of them is in the interior and an adjacent one, in $r$ or $\phi$, is in the atmosphere.  Radial and azimuthal boundary fluxes are uncoupled with the adjacent atmospheric cells because, in effect, $\Lambda$ is considered to act only in the vertical direction.  Therefore, the values of the $r$ and $\phi$ boundary fluxes are made equal to the vertical boundary fluxes under the assumption that the atmosphere surrounding an interior column has about the same properties in all other directions as it does in $z$.  Hence, a boundary cell will radiate (or gain) energy through a face contiguous with an atmosphere cell in the same manner as the column's vertical boundary.  This crude stair-step model for the disk surface enlarges the surface area and probably errs in the direction of enhanced disk cooling rates.

Finally, for the interior of the disk, the flow of energy is calculated by flux-limited diffusion, i.e.,
\begin{equation}{\rm F}_x = -\frac{16\sigma\beta T^3}{\chi_{\rm Ross} \rho}
\frac{\Delta T}{\Delta x} \label{eq11}\end{equation}
for radiative flux F$_x$ in direction $x=\left(r, \phi, z\right)$ and $\Delta x=\left( \Delta r,
\Delta \phi, \Delta z\right)$ on a cylindrical grid.  Because fluxes are measured at cell faces,
$\chi_{\rm Ross}$, $\rho$, and $T$ in equation (11) are the averages of the two adjacent cells that share the face at which the flux is being calculated.  The quantity $\beta$ is the flux limiter that, according to 
\citet{bodenheimer1990}, has the form
\begin{equation}\beta=\frac{2+y}{6+3 y + y^2}, \label{eq12}\end{equation}
for
\begin{equation}y=\frac{4}{\chi_{\rm Ross}\rho T}\left| \frac{\Delta T}{\Delta x}\right|.\label{eq13}
\end{equation}
The flux limiter is 1/3 in the high optical depth limit, while, in the optically thin limit, $\beta$ asymptotes to 1/$y$, making F$_x\rightarrow 4\sigma T^4$.  Remember, however, that the diffusion approximation is only used up to the boundary cell where $\tau_{\rm Ross}\gtrsim 2/3$.

Equation (11) is used to calculate all fluxes in the interior of the disk with four exceptions.  (1) When an adjacent cell is in the atmosphere, the flux at the common face between both cells is set to F$_{\rm bdry}$ as discussed above.  (2) If the atmosphere extends to the midplane, F$_{\rm bdry}=0$. (3) In the case that the interior extends all the way to the top of the grid, the flux at the top face is given by equation (8).  (4) All vertical fluxes at the equatorial plane are 0 because of the mirror symmetry assumption.  In the disk interior, we use the flux determined by equation (11) in equation (5) to calculate the divergence of the flux; for this reason, we refer to the cooling in the disk interior simply as  $\divflux $ cooling.  Appendix B demonstrates the accuracy of this algorithm.

Because the local radiative time scale, as defined by
\alphaeqn
\begin{equation}
t_{\rm \Lambda}{\rm ~or~}t_{{\nabla}\cdot {\rm F}}=
\frac{\epsilon}{\Lambda}\mathrm{~or~}\frac{\epsilon}{\divflux},
\label{eq14a} \end{equation}
can be much shorter than the Courant condition, we limit the local cooling time to 
be no less than about 10\% of the initial outer rotation period (ORP) of the gas disk at about 33 AU, where  1 ORP $\approx250$ yr.  Without this limiting, saw-toothing in the temperature structure of the disk becomes noticeable, and the $\Delta t$ becomes small and computationally prohibitive.  For similar reasons, the local heating time scale by artificial viscosity 
\begin{equation}
t_{\rm AV}= \frac{\epsilon}{\Gamma_{\rm AV}}\label{eq14b} \end{equation}
\reseteqn
 is also limited to be no less than about 0.1 ORPs.  Preliminary tests without these limits show that cooling and viscous heating times rarely become less than 0.1 ORPs when the disk has evolved away from its initial state and the cooling/heating-limited cells tend to be confined to uninteresting parts of the calculation. These local time scale limiters will not prevent fragmentation, because fragmentation occurs for $t_{\rm cool}\approx P_{\rm rot}$ for a $\gamma=5/3$ disk \citep{gammie2001,rice2003,mejia2005,mnras364l56}. 

\section{THE SIMULATION}

\subsection{Initial Conditions}

The initial model is the same one used for the CCT simulations described in \S 3.1 of Paper II, i.e., a nearly Keplerian, equilibrium disk of 0.07 M$_{\odot}$ that extends from 2.3 to 40.0 AU and orbits a star of 0.5 M$_{\odot}$.  The initial surface density profile is $\Sigma\sim r^{-0.5}$ and the initial equatorial temperature profile is $T\sim r^{-1}$ over most of the radial range. Opacities from \citet{dalessio2001} are used to compute $\chi_{\rm Ross}$ and $\kappa_{\rm Planck}$ (see Appendix A).  Temperatures are calculated every time step by assuming an ideal gas law, with densities and pressures already calculated in the code, and by iteratively solving for the molecular weight \citep[table taken from][]{dalessio1996}, because it too is a function of pressure or density and temperature.  Due to an error with the inclusion of helium in the solar mix, the mean molecular weight used for the temperature and pressure ranges in the simulation is around 2.7 when it should be close to 2.3.  This introduces a systematic offset no larger than about 16\% into the temperature, which directly affects the cooling rates.  However, as discussed below, the opacity law is roughly quadratic in temperature, which means the flux calculated from equation (11) is roughly quadratic in temperature, too. The error in the mean molecular weight should be an error that typically enhances the cooling by $1.16^2$ or about 4/3.  Because we expect the cooling to be enhanced, fragmentation should be more likely in the RC disk than it would be with the correct mean molecular weights.

The minimum allowed temperature at all times is 3 K to simulate radiating into empty space.  The temperature of this disk never reaches the dust sublimation temperature, near $T\sim 1400$ K \citep{muzerolle2003}, so the opacity is mostly due to dust. A maximum grain size $a_{\rm max} = 1\mu$m (see Appendix A) is used.

The main simulation presented in this paper evolves for a little over 16 ORPs, or about 4,000 yr.  The resolution used is the same as all the constant cooling time simulations of Paper II, but the $z$ direction is fixed at 32 zones.  The initial disk extends almost to the edge of the starting grid, namely $(r, \phi, z) = (256,128,32)$.  The $r,z$ cross-section of this disk in the initial grid is accurately portrayed in Figure 1.  The grid is extended to 512 in $r$ when the disk expands after the first few ORPs.  Two phases of the simulation are also run in a high azimuthal resolution grid, $(r, \phi, z)=(512,512,32)$ to test for fragmentation (see \S 5.2).   A random cell-to-cell density perturbation of amplitude $\left| \Delta\rho/\rho\right|=10^{-4}$ is applied at the very first step of the run, which allows spiral modes to grow from the background noise as the disk cools. 

\subsection{The Evolution}

Figure 2 shows number density contours in the equatorial plane of the RC simulation.  The RC disk undergoes the same four evolutionary phases described for CCT in Paper II, namely the axisymmetric, burst, adjustment, and asymptotic phases.  In the axisymmetric phase, the disk shrinks slightly while cooling dominates over AV heating.  The outer disk produces a dense, transient ring, and it becomes part of the spiral arms once instability sets in. The spiral structure develops between 2 and 3 ORPs of evolution, and the disk nearly doubles in size during the burst phase.  The first expansion happens right at 4 ORPs, but, during adjustment, the disk clearly oscillates and reaches comparable sizes at 6, 7.7, and 10 ORPs.  Subsequently, in the asymptotic phase, the disk maintains roughly the same outer radius. Often, one long spiral arm dominates the structure, and the disk has a more lopsided appearance than the CCT disk's analyzed in previous papers. Because the central star is kept fixed at the grid center, the dynamics of the these one-armed structures may not be accurately treated.  We are developing routines to relax this constraint in the future by calculating the star's motion explicitly.  Methods like moving the star to a location that brings the center of mass to the center of the grid \citep{boss1998} are not employed because this might incorrectly treat the dynamics as well. Table 1 lists several properties of the initial and final disk at the equatorial plane for 10, 30, and 50 AU in radius.

\begin{figure}[ht!]
\begin{center}
\includegraphics[width=12cm]{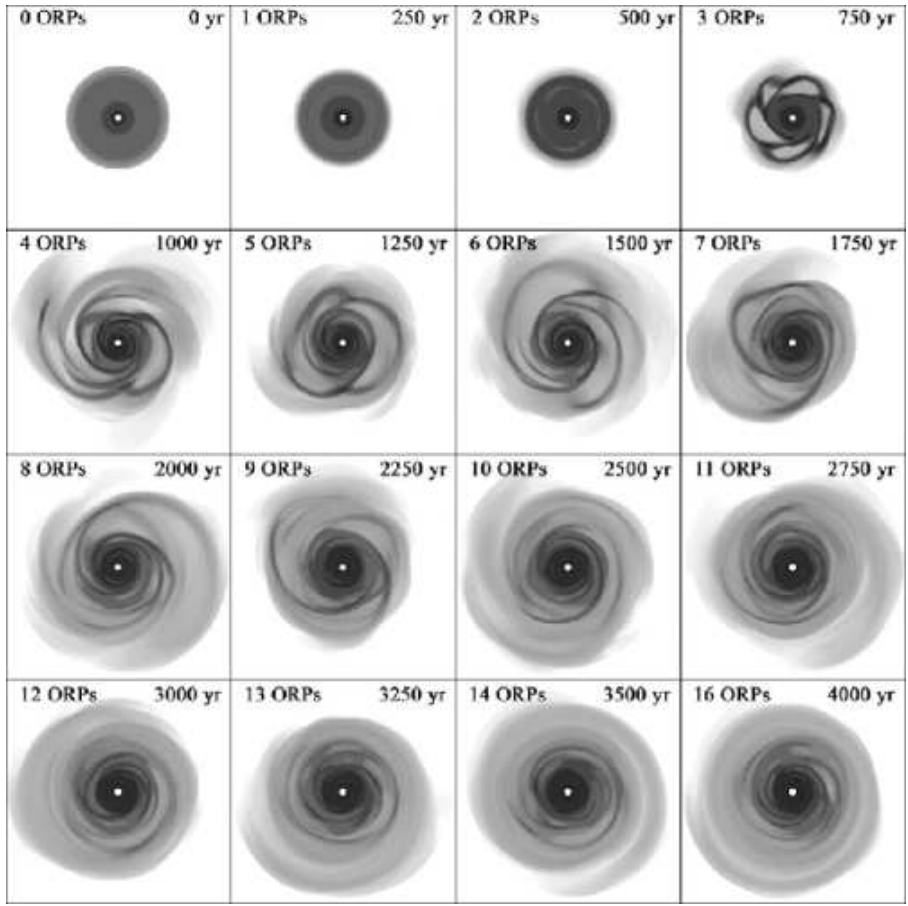}
\caption{Evolution of the RC disk.  All the images show number densities at the equatorial plane on a logarithmic color scale that stretches four orders of magnitude with red representing the highest density and blue the lowest.  The squares enclose the disk at maximum size, 170 AU on the side. A movie of the simulation is available at {\it http://westworld.astro.indiana.edu/} under the {\it Movies} link. [Journal version: Evolution of the RC disk.  All the images show number densities at the equatorial plane on a logarithmic gray scale that stretches four orders of magnitude with black representing the highest density and white the lowest.  The squares enclose the disk at maximum size, 170 AU on the side. A movie of the simulation is available at {\it http://westworld.astro.indiana.edu/} under the {\it Movies} link.]}
\label{f2}
\end{center}
\end{figure}

\begin{table}[ht]
\caption{Characteristics of the RC disk at 0 and 16 ORPs.  Here $R_{\rm disk}$ is the outer radius of the disk.}
\begin{center}
\begin{tabular}{|c|c|c|c|c|c|c|c|}\hline
$t$(ORP) & $R_{\rm disk}$ & Grid size $(r, \phi, z)$ & $r$ (AU) & $T$ (K) & $Q$ & $n(\rm cm^{-3})$
    & $ \Sigma \rm (g~cm^{-2})$\\ \hline\hline
 & & & 10 & 102.5 & 7.4 & 1.5(12) & 207.9 \\   \cline{4-8}
0 & 40 & 256,128,32 & 30 & 45.9 & 1.6 & 4.5(11) & 141.6 \\ \cline{4-8}
 & & & 50 & ... & ... & ... & ... \\ \hline
 & & & 10 & 24.3 & 1.8 & 8.5(12) & 457.8\\ \cline{4-8}
 16 & 65 & 512,128,32 & 30 & 8.4 & 1.5 & 5.2(11) & 49.8 \\ \cline{4-8}
 & & & 50 & 3.4 & 2.5 & 4.8(10) & 11.9\\ \hline
 \end{tabular}
\end{center}
\label{t1}
\end{table}

\subsection{Mass and Density Distribution}

As expected, the structure of the disk changes significantly once the gravitational instabilities develop.  Large amounts of mass move radially as the spiral pattern becomes nonlinear.  We calculate average mass fluxes by differencing the total mass inside cylinders at two separate times and dividing by that time interval.  This method should represent the average mass fluxes in the disk as calculated by the code's second-order flux scheme.  Post-analysis methods for calculating the $\dot M$'s prove to be unreliable because the fluctuations in the instantaneous mass fluxes are much larger by a factor of a few to about 10 than the net mass flux and are highly variable.  As discussed in \S 5.1, the picture of mass slowly diffusing between different locations is incorrect for the RC disk.  Between 2 and 4 ORPs, most of the net inward mass transport happens between about 9 and 27.7 AU, while net outward transport occurs outside 27.7 AU (Fig.\ 3).  The accretion rates inside 7 AU are negligible.  The peak transport rates are $1.75\times 10^{-5}$ and $3.25\times 10^{-5}$ M$_{\odot}\rm/yr$ inward and outwards, respectively.  During the adjustment phase, the mass transport decreases by about an order of magnitude with typical rates of about $10^{-6}$ M$_{\odot}\rm/yr$ and is mostly inward over 10 to 40 AU.  After 10 ORPs, which marks the beginning of the asymptotic phase, the inflow rate peaks at $8.5\times 10^{-7}$ M$_{\odot}\rm/yr$, averages about $2\times 10^{-7}$ M$_{\odot}\rm/yr$, and is mostly inward over 10-25 AU.

\begin{figure}[ht!]
\begin{center}
\includegraphics[width=10cm]{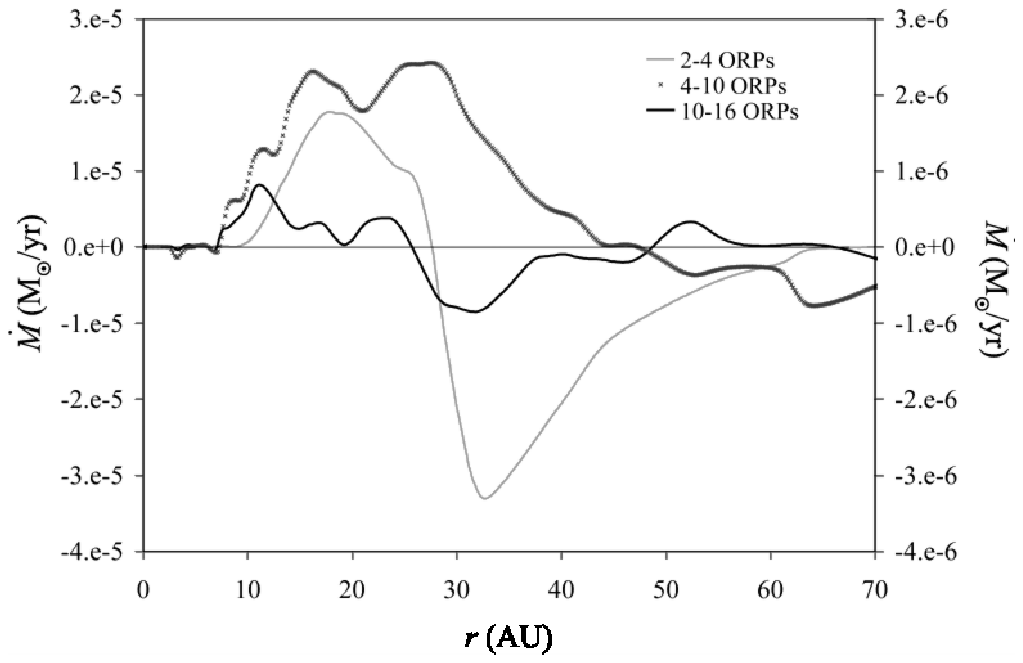}
\includegraphics[width=8cm]{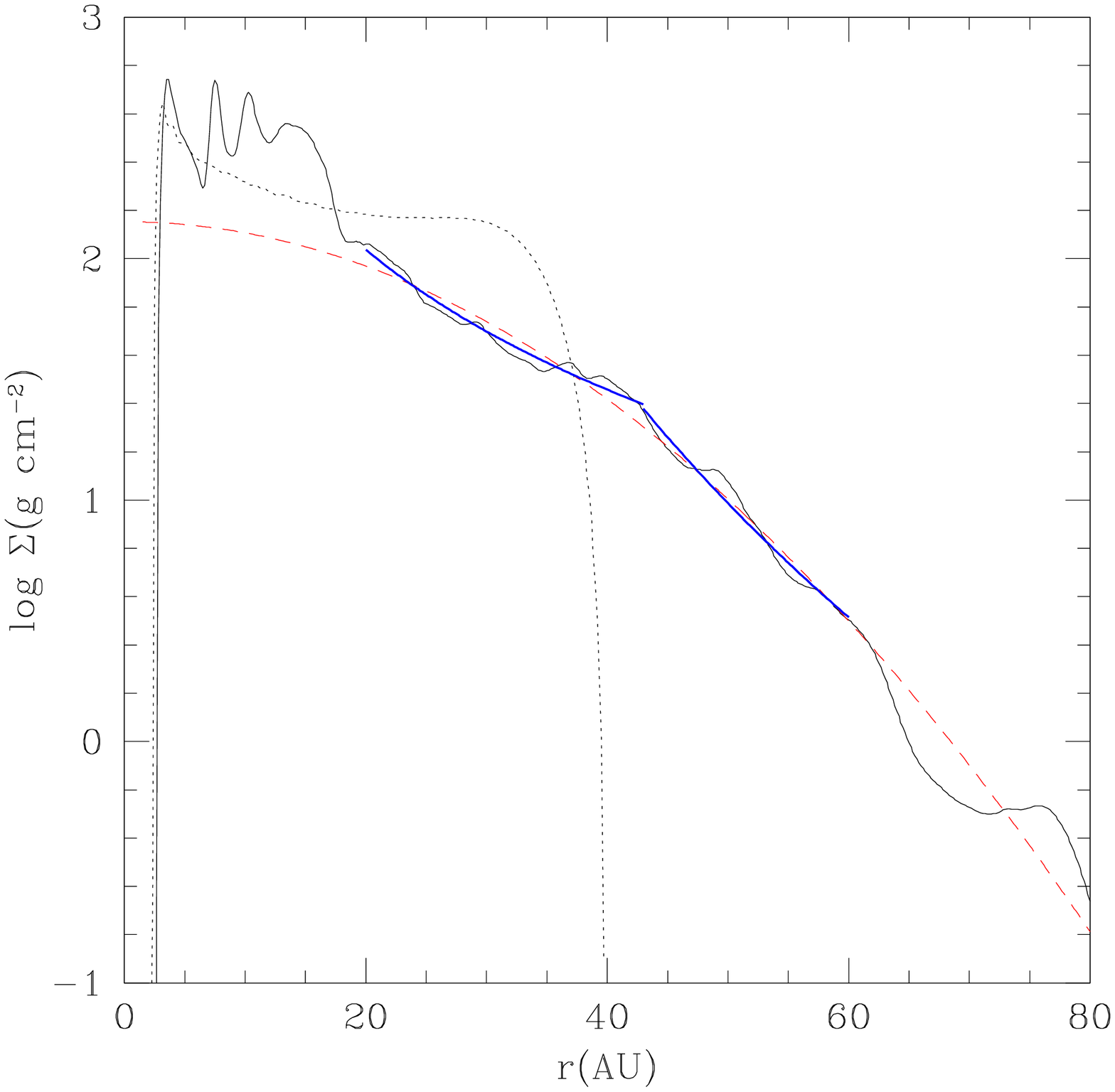}
\caption{Global mass redistribution.  The top panel shows the average mass transport rate calculated by differences in the total mass fraction as a function of radius for three different times: between 2 and 4 ORPs, 4 and 10 ORPs, and 10 and 16 ORPs.  The red [gray] curve is scaled to the left ordinate while the blue [starred] and green [dark] curves are scaled to the right ordinate.  The bottom panel shows the surface density as a function of radius for the initial disk (dotted) and the final state (solid), which reflects the density profile during the asymptotic state.  The dashed curve shows that the density profile for $r\gtrsim 20$ AU follows a Gaussian distribution as described in the text.  However, the surface density profile can also be broken down into two power laws (blue [dark] curves).}
\label{f3}
\end{center}
\end{figure}

Figure 3 also illustrates how the $r^{-0.5}$ surface density profile of the initial disk is lost, and at 4 ORPs the disk approximately achieves its final distribution with the profile obeying a Gaussian of the form $\Sigma = \Sigma_{\circ} 10^{-\left( r/r_e\right)^2}$ (dashed curve). The least-squares fit in log-linear space with the dummy variable $x=r^2$ yields $r_e=46.7$ AU when fit between $r=[20,60]$ AU.  However,  on the intervals $r=[20,43]$ AU and $r=[43,60]$ AU the surface density profile seems to follow two different power laws $\Sigma\sim r^{-\nu}$.  For the inner interval, $\nu=-1.93$, and for the outer interval, $\nu=-5.97$.  The Spearman correlation coefficients are $R=-0.992,-0.986,-0.985$ for the exponential, the inner power law, and the outer power law, respectively.  Michael et al. (2006, in preparation) explore the effects of initial conditions on these final surface density distributions.  

The disk inside 20 AU is not well-described by any simple monotonic function.  There is a ring at 7.5 AU, and another seems to be forming at 10.5 AU, both resembling those that appeared in the CCT simulations.  These rings are obvious in the cylindrical mass plot (Fig.\ 4).  The broad peak seen around 15 AU in the cylindrical mass plot is not a true ring but represents mass concentrations in spirals that come together in the 10.5 AU ring.  The mass of the 7.5 AU ring continuously increases and reaches 5 Jupiter masses (M$_{\rm J}$) at the end of the simulation.  The 10.5 AU ring has a mass of 9.0 M$_{\rm J}$ at 16 ORPs.  See \S 5.1 for more details and a discussion about the causes and possible significance of these rings.  

\begin{figure}[ht!]
\begin{center}
\includegraphics[width=12cm]{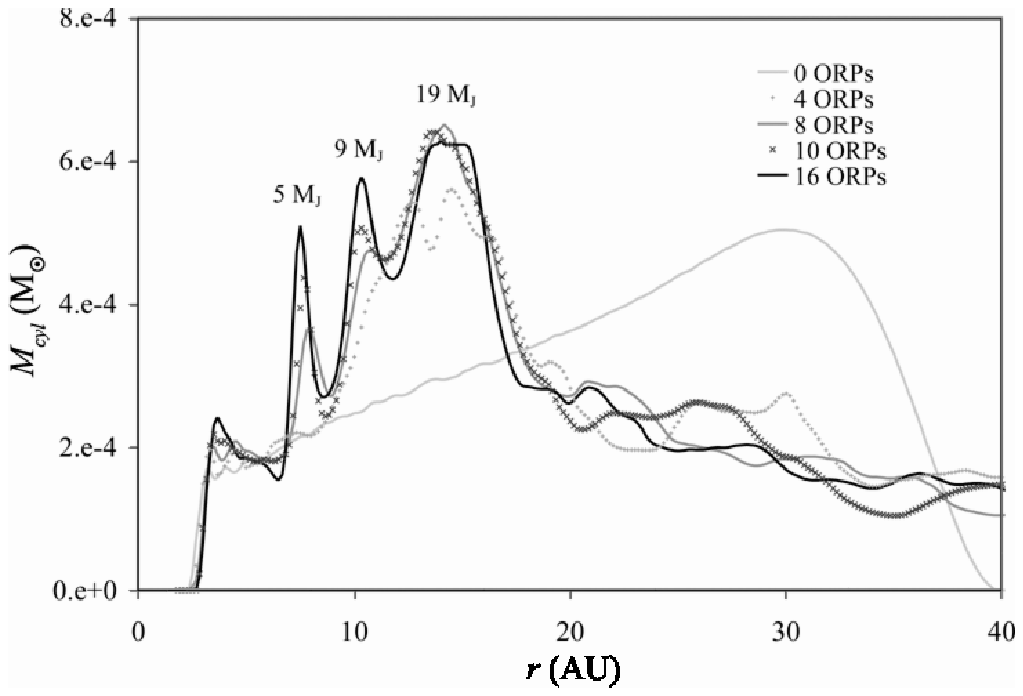} 
\caption{Rings in the inner disk. This plot shows the total mass per grid cylinder (1/6 AU in width) at various times during the evolution.  The rings are labeled with their respective masses. }
\label{f4}
\end{center}
\end{figure}

\subsection{Heating and Cooling}

The top panel of Figure 5 shows the total internal energy $U$ contained in the disk as a function of time. The same plot also tallies total energy losses due to cooling in the optically thin regions (atmosphere) and in the interior of the disk as well as the total energy gained from heating by shocks.  The overall behavior of the total internal energy of this disk is similar to the CCT disk's behavior.  It must be noted that the $-P{\bf \del}\cdot{\bf v}$ term is not included in Figure 5.  Therefore, just adding the energy gains and subtracting the energy losses due to $\Lambda$, $\Gamma$, and $\divflux$ from the initial value of the internal energy will not result in the final value of $U$.  Moreover, due to coarse resolution, some of the $\tau < 2/3$ emission at the boundary of the atmosphere and disk interior is effectively included in the $\divflux$ curve. 

\begin{figure}[ht!]
\begin{center}
\includegraphics[width=9cm]{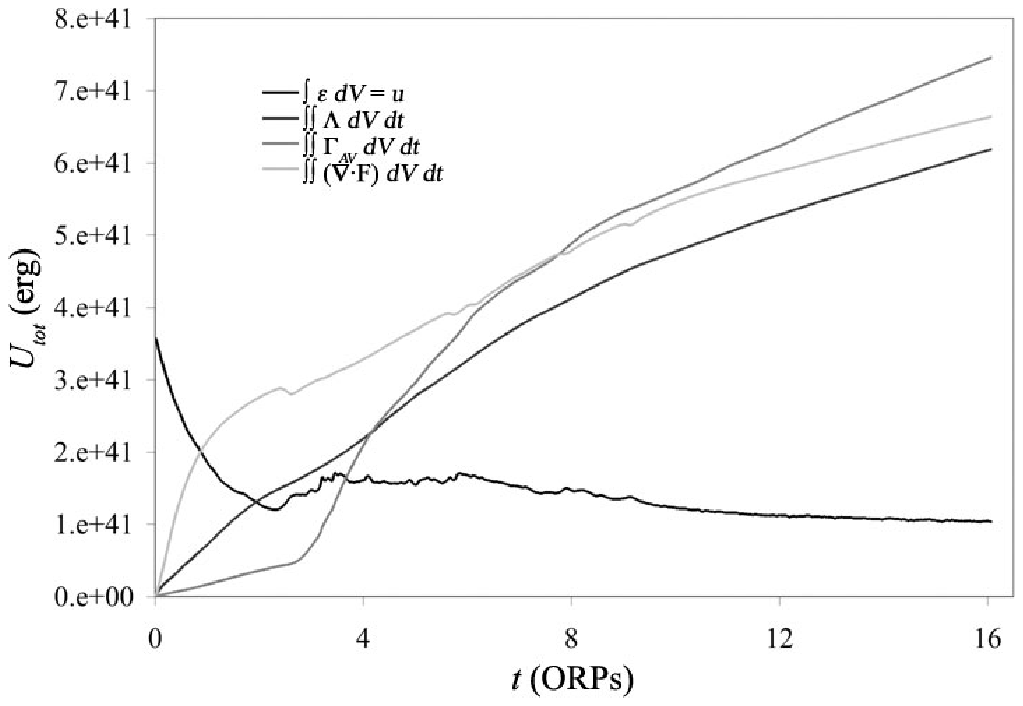} \includegraphics[width=9cm]{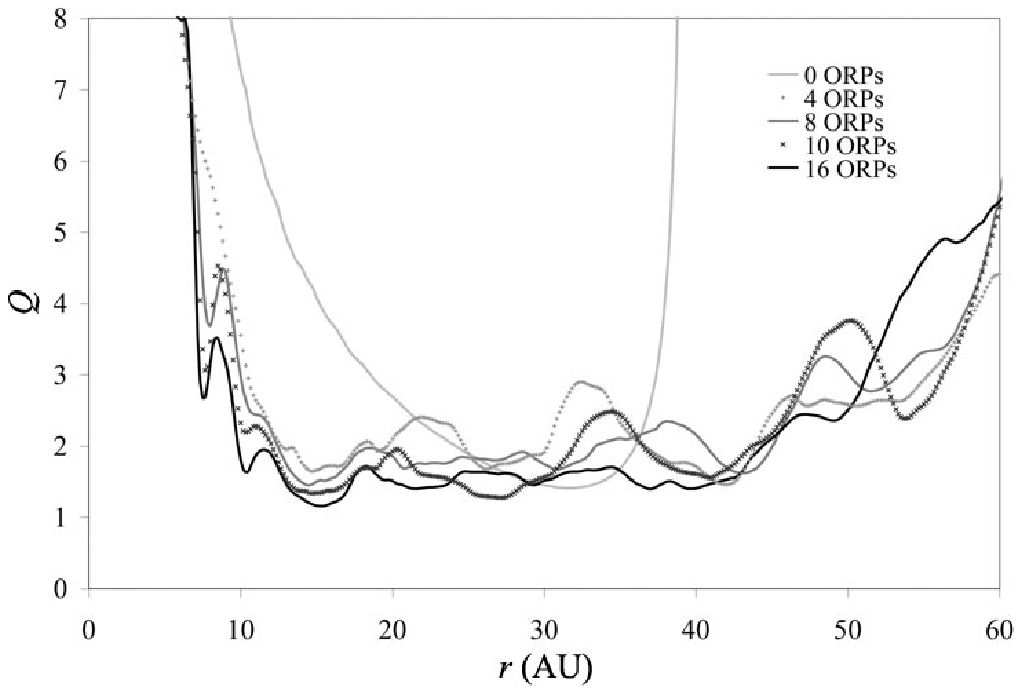} 
\caption{Total energies (top) and Toomre $Q$ (bottom) as a function of time.  The green curve in the top panel shows the total internal energy in ergs.  The other curves show the total cumulative energy gains and losses due to the cooling and heating processes, also in ergs.  Plotted separately are the contributions by shock heating, diffusion in the disk interior, and radiative cooling in the atmosphere.  The bottom panel shows the Toomre $Q$ every few ORPs.  }
\label{f5}
\end{center}
\end{figure}

The evolution of the Toomre $Q(r)$ of the disk is shown in the bottom panel of Figure 5.  As in Paper II, $Q$ is calculated using the full vertical (top + bottom) surface density and both the sound speed and the angular speed evaluated at the midplane, where $\Omega$ replaces the epicyclic frequency in equation (1) for a nearly Keplerian disk.  All these quantities are azimuthally averaged before $Q$ is computed.  At 16 ORPs, the average value between about 14 and 43 AU is 1.5, with a standard deviation $\sigma$ of 0.14.
Together, the internal energy curve and the Toomre $Q$ curves suggest that, after about 10 ORPs, RC is in an asymptotic-like phase as seen in the CCT disk with the disk remaining marginally unstable and shock heating roughly balancing cooling. The distinction asymptotic-{\it like} is made because RC's cooling times continuously adjust, and its evolution is different from the CCT disk's in that disk properties change noticeably even in this phase.   We estimate the evolution time $t_{\rm ev}$ for the disk in the asymptotic phase by $t_{\rm ev}\sim U / \dot{U}$ and by $t_{\rm ev} \sim M_{\rm in}/\dot{M}$, where $M_{\rm in}$ is the mass contained between 20 and 26 AU, {\it ceteris paribus}.  Both methods yield $t_{\rm ev} \sim $ few $\times 10^{4}$ yr; according to these estimates, the asymptotic phase is short-lived compared with typical total disk lifetimes \citep[e.g.,][]{hartmann_asp341}. However, effects like grain growth would significantly alter the calculation.  Results presented by \citet{cai_letter_2006} indicate that grain growth in a disk where the dust remains well mixed can lead to increased cooling times, which lead to weaker GIs and to a slower evolution.  

The average effective temperature (Fig.\ 6) over the 10-16 ORP time interval, computed by taking the fourth root of the azimuthally averaged $T_{\rm eff}^4$, seems to follow a temperature profile $T= T_{\circ}10^{- r/r_e}$.  For RC, $r_e=44.8$ AU when fit over the region $r=(3,50]$ AU; this result is fairly insensitive to the limits chosen for the fit and has a Spearman correlation coefficient $R=-0.988$ in the log-linear space. However, if we choose to fit only the region $r=(3,20]$ AU, a power law $T\sim r^{-0.59}$ seems to approximate the profile well with an $R=-0.957$.  This value of the exponent is primarily due to the hot inner disk boundary between 2.3 and 5 AU.

\begin{figure}[ht!]
\begin{center}
\includegraphics[width=8cm]{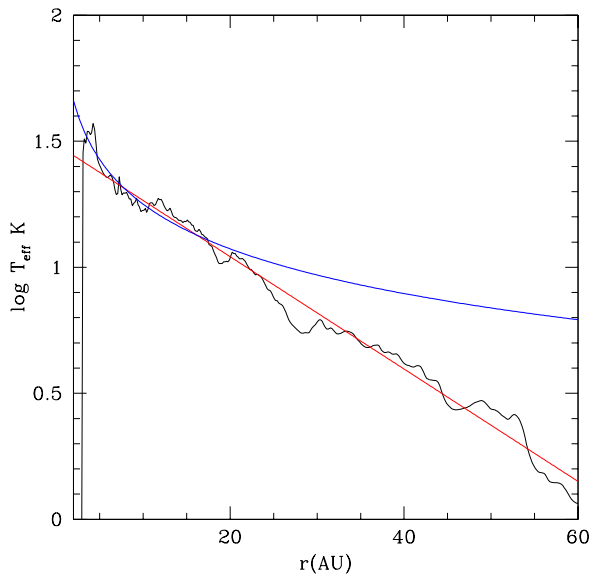} 
\caption{Average effective temperature vs.\ radius time averaged over 10-16 ORPs.  The profile that best fits most of the disk is an exponential, $T= T_{\circ}10^{- r/r_e}$ for  $r_e=44.8$ AU, delineated by the red [gray] line.  However, the $r=(3,20]$ AU region appears to obey a power law $T\sim r^{-0.6}$ shown by the blue [dark] curve.}
\label{f6}
\end{center}
\end{figure}

The series of side-view maps shown in Figure 7 illustrate density and temperature at the end of the run.  The left column shows a vertical cut through the disk, while the right column shows the azimuthal average.  Generally speaking, the disk is dense in the inner parts and is more diffuse and  slightly flared in the outer disk.  Most of the disk interior, i.e., the region inside the disk photosphere, is contained inside about 25 AU, and its half thickness is only about 1 AU in the vertical direction, shrinking considerably from the interior of the initial disk with a radial extent of 37 AU and a vertical half-thickness of 3 AU.  This is also the region where the temperatures remain higher than 10 K.  The rest of the disk has cooled to temperatures of only a few Kelvin.  However, there are streaks of higher temperatures in the outer disk created by shocks.  Shock heating, in general, has the shortest times in the upper layers \citep{pickett2000} when compared to the local orbital time, and this heating is probably due to shock bores, which are the nonlinear outcome of spiral shocks in disks \citep{boleyshockbores}.  Typical cooling/heating times are a few to tens of ORPs, which indicates that the local cooling limiters defined in equation (14a) do not hinder the evolution of the disk in the asymptotic phase.

\begin{figure}[ht!]
\begin{center}
\includegraphics[width=16cm]{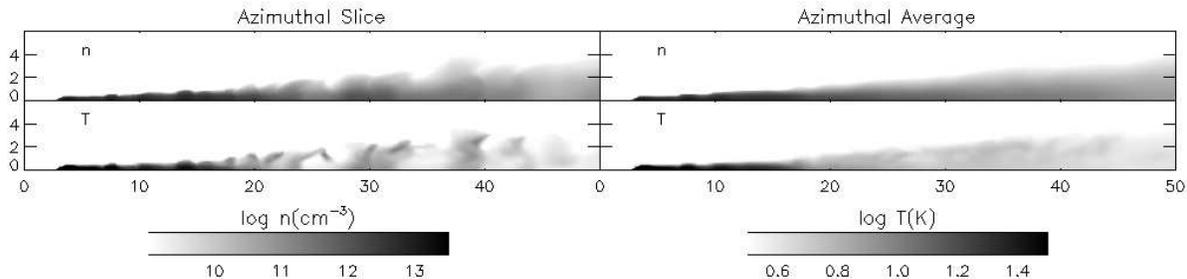} 
\caption{Vertical structure of the disk at 16 ORPs for one slice of the grid (left) and the azimuthal average of all the slices (right).  Number density is represented in the top panels and temperature in the bottom.}
\label{f7}
\end{center}
\end{figure}

\section{RADIATIVE COOLING VS.\ CONSTANT COOLING TIME}

\begin{figure}[ht!]
\begin{center}
\includegraphics[width=12cm]{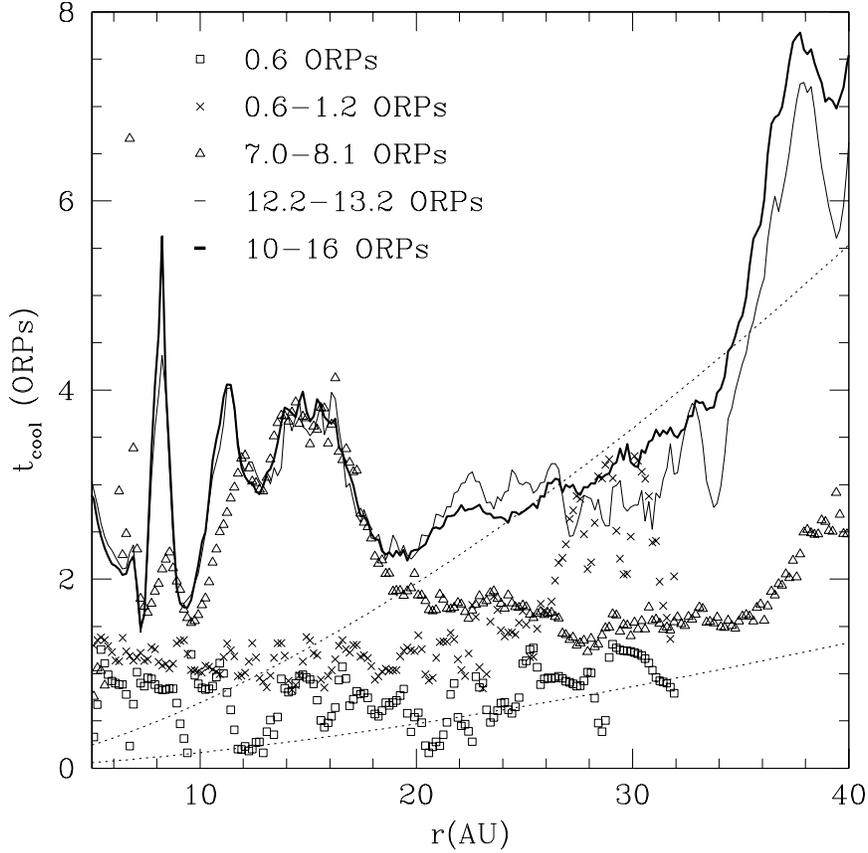} 
\caption{Azimuthally averaged column-wise cooling times as a function of disk radius for a snap shot at 0.6 ORPs (squares) and for the temporal averages between 0.6 and 1.9 ORPs (crosses), 7 and 8.1 ORPs (triangles), and 12.2 and 13.2 ORPs (light curve).  The heavy curve profile represents the average cooling times between 10 and 16 ORPs as a function of $r$.   For the profiles that continue past 40 AU, the cooling times continue to increase and become well over 100 ORPs.  The profiles for 0.6 and 0.6-1.9 ORPs are radially smoothed.  The dotted curves represent $t_{\rm cool}=6/\Omega$ (lower) and $t_{\rm cool}=25/\Omega$ (upper).}
\label{f8}
\end{center}
\end{figure}

The RC disk undergoes gravitational instability much sooner than the $t_{\rm cool} = 2$ ORP CCT disk in Paper II.  During the axisymmetric phase, the typical cooling time of the optically thick ($\tau_{\rm Ross} > 2/3$) interior disk is shorter than 2 ORPs, so it cools to instability in about half the time.  Short initial cooling times are due to strong departures from uniform radiative equilibrium in the initial model.  As the simulation progresses and the disk adjusts, the typical cooling times increase, and, at the end, an entire column radiates its internal energy in a matter of a few to several tens of ORPs.  We compute the azimuthally averaged column-wise cooling time by $t_{\rm cool} = \left< \varepsilon_{\rm col} \right>/ \left<\lambda\right>$, where $\varepsilon_{\rm col}=\int_0^\infty\epsilon~{\rm d}z$ is the vertically integrated internal energy of the half disk in a given column and $\lambda=\int_0^\infty \left( \divflux + \Lambda \right)~{\rm d}z$.  The brackets indicate azimuthally averaged quantities.  As shown in Figure 8, the azimuthally averaged column-wise cooling times generally increase with time and with disk radius, which is consistent with the argument that $t_{\rm cool}\sim T^{-3}\tau$ for large $\tau$ and $t_{\rm cool}\sim T^{-3} \kappa_{\rm Planck}^{-1}$ for small $\tau$ by \citet{durisen_ppv_chapter} \citep[see also][]{rafikov2005}.  Also shown in Figure 8 are curves for $t_{\rm cool} = 6/\Omega$ and $t_{\rm cool}= 25/ \Omega$.  Once the disk moves away from its initial state, the cooling times never drop below the \citet{mnras364l56} fragmentation limit ($t_{\rm cool} = 6/\Omega$).  We discuss the significance of the $t_{\rm cool}=25/\Omega$ curve in \S 5.1.2.

\begin{figure}[ht!]
\begin{center}
\includegraphics[width=16cm]{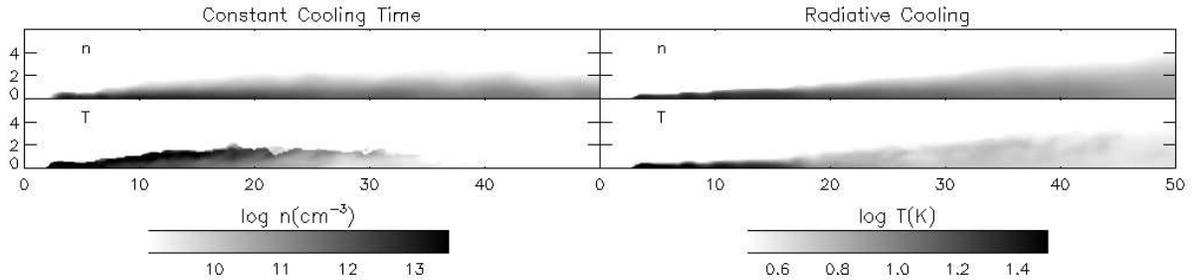} 
\caption{Comparison of the azimuthally averaged vertical structure of the RC and CCT disks at 16 and 23.5 ORPs, respectively. CCT's surface is much hotter than RC's because the cooling time scale is spatially uniform, while heating by shocks is highly nonuniform.}
\label{f9}
\end{center}
\end{figure}

Figure 9 shows the final states of RC and CCT at 16 and 23.5 ORPs, respectively. We choose to compare the disks with each other at 16 and 23.5 ORPs because, although  CCT has been evolved for a much longer time period, RC evolves much more quickly. The high temperatures in the upper layers of CCT are due to heat deposited by shocks, and, because the cooling time is constant everywhere, the upper layers cool more slowly there than they do in the RC simulation.  This illustrates the importance of shock bores in disks.  \citet{boleyshockbores} demonstrate that shock bores can deposit large amounts of energy into the disk atmosphere and its middle layers.  These high-altitude temperatures should make the disk convectively stable (see \S 5.3); however, since the energy is deposited in layers that can cool efficiently, shock bores could effectively keep a disk cooler overall by depositing energy from the post-shock region of spiral waves into the high-altitude gaseous layers.

The instabilities are stronger in CCT, with an average Toomre $Q$ of 1.44 ($\sigma  = 0.24$) compared with 1.50 ($\sigma = 0.14$) for RC, each measured between 14 and 43 AU.  To measure the nonaxisymmetric structure in both disks, as done in \citet{cai_letter_2006}, we compute the time-averaged, integrated Fourier amplitudes $\left< A_m\right>$, where we define 
\begin{equation} A_m=\frac{ \int \rho_m~r{\rm d}r{\rm d}z }{ \int \rho_0~ r {\rm d}r{\rm d}z};\label{eq15}\end{equation}
$\rho_{0}$ is the axisymmetric component of the density and $\rho_m$ is the total Fourier amplitude of the $\cos m \phi$ and the $\sin m \phi$ density component \citep[see][]{imamura2000}. Summed over $m=2$ to 63, the integrated Fourier amplitude for $t_{\rm cool}=2$ ORP CCT is about 2.7 when averaged between 21.4 and 23.4 ORPs, whereas the amplitude is about 1.6 for RC when averaged between 14 and 16 ORPs, which is consistent with CCT having stronger GI activity than RC in the asymptotic phase.  This is expected given the substantially longer column-wise cooling times in RC.

\section{DISCUSSION}

\subsection{Mass Transport and its Locality}

A principal question
 \citep{pringle1981,laughlin_rozyczka1996,balbus_papaloizou1999, gammie2001, lodato_rice2004,mejia2005} 
is the following: Can mass and angular momentum transport in  GI-active disks be modeled by a local $\alpha$ prescription, or do the long-range torques that GIs produce make such an approximation misleading?  If GIs have a global effect on mass transport in the disk, substructure caused by spatially and temporally variable accretion rates, such as rings or density enhancements, can be missed by assuming that the disk evolves like an $\alpha$-disk.  In this section, we quantify the effects of GIs on mass transport in the RC disk and compare that behavior with CCT's to address the locality of GIs in disks.

\subsubsection{Torques and Modes}

GIs redistribute mass and transport angular momentum over most of the RC disk chiefly through the torque exerted on the disk by the spiral waves.  Peak mass transfer rates are larger than $10^{-5}$ M$_{\odot}$/yr during the initial burst and asymptote to a few $\times 10^{-7}$ M$_{\odot}$/yr at later times, comparable with those of the CCT disk.   To investigate mass and angular momentum transport in RC further, we compute the torque in the disk.  Following \citet{lynden-bell_kahlnajs1972}, one may write the torque ${\bf C}$ due to the outer disk on the disk inward of some distance ${\bf x}$ from the origin by integrating the gravitational stress tensor $\widetilde{T}^{\rm grav} $ and the Reynolds stress tensor $\widetilde{T}^{\rm Reyn} $ over the surface of a cylinder;
\begin{equation} {\bf C}=\int {\bf x}\times\widetilde{T}^{\rm grav}\cdot{\rm d}{\bf S} +
\int {\bf x}\times\widetilde{T}^{\rm Reyn}\cdot{\rm d}{\bf S}.\label{eq16}\end{equation}
For the gravitational contribution, instead of using the stress tensor directly, one may integrate over the total volume by
\begin{equation} {\bf C}^{\rm grav} = -\int \rho~{\bf x}\times {\bf \del}\Phi~{\rm d}V,\label{eq17}\end{equation}
where $\Phi$ is the gravitational potential.  Furthermore, we are only interested in the $z$ direction of the torque $C_z$ given by
\begin{equation} C_z^{\rm grav} = -\int \rho\frac{\partial \Phi}{\partial \phi}~{\rm d}V.\label{eq18}\end{equation}
The contribution to the $z$-torque from the Reynolds stress is calculated by
\begin{equation} C_z^{\rm Reyn}=\int\int \rho\delta v_r \delta v_{\phi}~r{\rm d}z{\rm d}\phi,\label{eq19}\end{equation}
where $\delta v_{\phi}$ is the fluctuating component of the azimuthal velocity and $\delta v_r$ is the fluctuating component of the radial velocity.  The fluctuating components are calculated by subtracting the mean velocity, found by azimuthally and vertically averaging the velocities with a density weighting at each $r$, from the velocity in each cell for the corresponding $r$.  A more detailed discussion of the analysis associated with equations (17) to (19) will be given in Michael et al.\ (in preparation).

	Figure 10 shows the time-averaged gravitational torque (heavy black) and gravitational plus Reynolds torque (light black) over 10-16 ORPs; the red curve is explained below.  Superposed onto the plot is the average $\dot{M}(r)$ over the same interval.  The ring at 10 AU (see Fig.\ 4) is formed where the total torque precipitously declines.  This also agrees with the sudden drop in accretion rate, as discussed in \S 3.3, near the same radius, and indicates that the formation of this ring can be understood as the place where mass piles up due to a diminution of GI torques at this radius. It is unclear whether this ring would survive if our code included additional accretion mechanisms in the inner disk, e.g., an $\alpha$ viscosity caused by the magnetorotational instability \citep{balbus_hawley1991,gammie1996}.  This is a subject for future study.

\begin{figure}[ht!]
\begin{center}
\includegraphics[width=12cm]{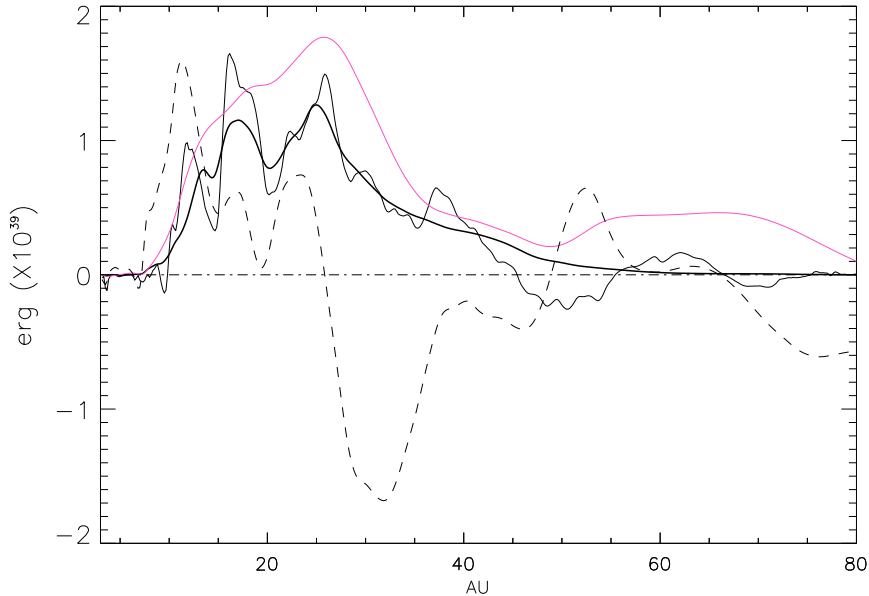}
\caption{Time-averaged gravitational and Reynolds torque profile (thin solid) and just the gravitational torque (heavy solid) for the 10-16 ORPs evolution time interval.  We use the sign convention that a positive value indicates outward angular momentum transport. Superposed on the plot, unscaled to the ordinate, is the time-averaged mass accretion rate (dashed line).  The major break between inflow and outflow occurs near the maximum of the gravitational torque ($r=$ 26 AU), but the highest mass inflow occurs inside the global maximum of the total torque at $r=16$ AU.  The red [gray] curve is the torque profile derived from the $\dot{M}$'s for the 10-16 ORP interval.  See section \S 5.1.3 for more detail.}
\label{f10}
\end{center}
\end{figure}

	To complement this figure, we present a periodogram (Fig.\ 11) for the $m=2$ Fourier component of the nonaxisymmetric density structure in RC as done in \S 3.3.2 of Paper II for CCT.  The periodogram detects Fourier component $m$-armed density fluctuations that have coherent pattern frequencies over a large range of radii.  The contours show power, where dark red is the strongest and light blue is the weakest, and the dark curves indicate corotation $(\Omega = \Omega_{\rm pattern})$ with the gas and the inner and outer Lindblad resonances $(\Omega = \Omega\pm\kappa_e/m)$.  The resonance curves are noisy in the ring radii range, even after some smoothing, because the rings create strong pressure gradients, which affect the epicyclic frequency.  There is significant power at corotation at a period of about 4/7 ORP near $r = 26$ AU.  This corresponds to the mass transport inflow/outflow boundary, which also corresponds to the global maximum in the gravitational torque and a local maximum in the total torque in Figure 10.  Moreover, there are swathes of power near periods of 1/2 and 1 ORP.  The 1/2 ORP swath has its outer Lindblad resonance at the inflow/outflow boundary, and the 1 ORP swath has its inner Lindblad resonance at the inflow/outflow boundary. The alignment of the three major swathes of power with corotation, inner Lindblad, and outer Lindblad resonances, combined with coincidences between corotations and significant $\dot{M}$'s, as shown in Figure 11, suggests dominance of low-order modes in the mass and angular momentum transport. In addition, there is power on the corotation curve at 1/9, 1/6, and 1/4 ORP; the rings are between these locations.  As suggested by \citet{durisen2005}, resonances may play a role in forming and maintaining the rings.  
	
The three major power swaths are also noticeable in the periodograms for $m=3$, 4, and 5, but they  become less prominent with increasing $m$.  Moreover, the periodograms become more noisy with increasing $m$ in the sense that there are many other swaths of power present.  Therefore, we present in Figure 12 an amplitude spectrum for the $A_m$'s defined in equation (15) for $m=[1,63]$.  Clearly, the low-order $m$'s have the largest amplitudes. The $m=1$ amplitude is probably larger than what it would be if we would allow the star respond to the background potential, and the extra power in the $m=1$ mode probably affects the disk dynamics in some respects but not in the gross behavior, which we believe based on the consistent results between the power in the $m=2$ mode, the periodograms, and the mass transport.   For $m=[2,63]$, the data are fit by the model curve
$A_m \sim (m^2 + m_0^2)^{-1.64}$, where $m_0=7.46$.  This curve highlights that $A_m$ asymptotes to a power law for large $m$ values.  Along with the noise in the periodograms, the $A_m$ profile suggests that GIs lead to a cascade of power among many modes resulting in {\it gravitoturbulence} \citep{gammie2001}.  In fact, the radial wavenumber power spectrum shown by \citet{gammie2001}, which excludes low-order modes because a shearing box approximation is used, appears to be consistent with our power spectrum for $m>10$.  Understanding the slope of this spectrum for large $m$ values and determining the uniqueness of this profile are topics for future study.

\subsubsection{Is Angular Momentum and Energy Dissipation Accurately Described by an $\alpha$ Prescription?}

\begin{figure}[ht!]
\begin{center}
\includegraphics[width=12cm]{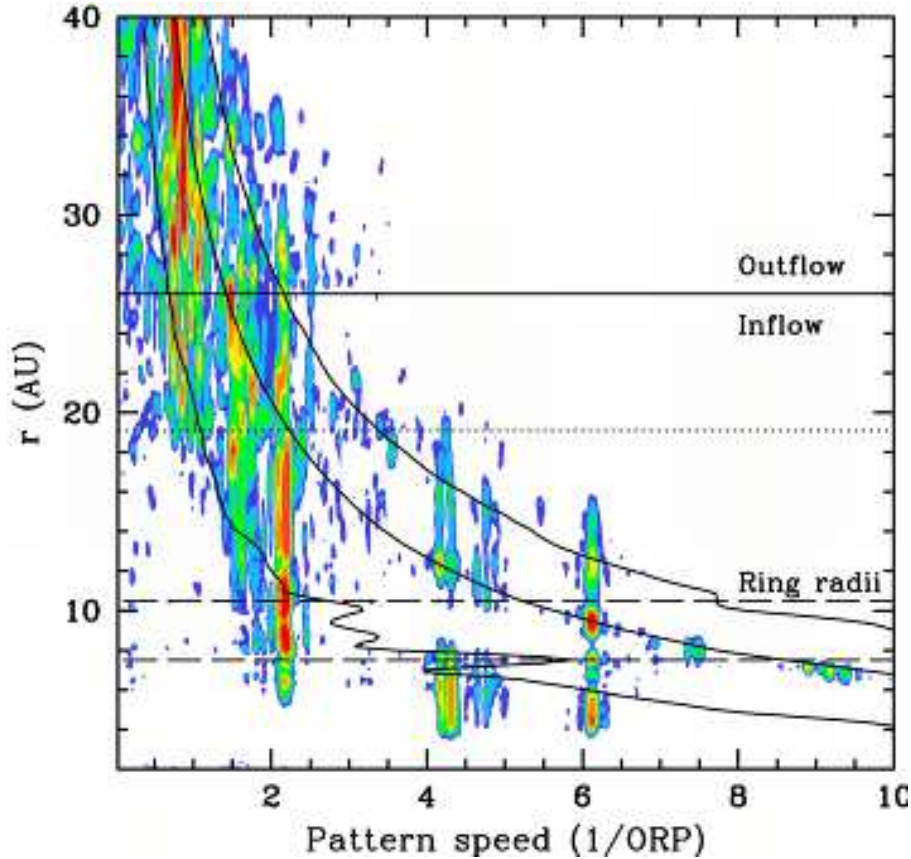}
\caption{Periodogram for the $m = 2$ mode.  Even through the noise, three large swathes of power are noticeable at about 1/2, 4/7, and 1 ORP, which cross the outer Lindblad, corotation, and inner Lindblad resonances, respectively, at the mass inflow/outflow boundary. Moreover, there is significant power at about 1/9, 1/6, and 1/4 ORP at corotation.  The radii corresponding to these local power maxima straddle the rings.  The dotted line indicates where the mass accretion rate almost returns to zero and corresponds to the corotation resonance for the swath of power near 1/2 ORP.  This line also represents where the exponential surface density profile abruptly ends with a sudden plateau in the overall profile and the formation of rings (Fig.\ 3).}
\label{f11}
\end{center}
\end{figure}

The torque can also be used to calculate a \citet{alphadisk} $\alpha$ for the disk.  Define
\begin{equation} \alpha = \left| \frac{ {\rm d} \ln\Omega} { {\rm d}\ln r}\right|^{-1}
\frac{ \left<G_{r\phi}\left> + \right<R_{r\phi}\right> }{\left<\Sigma c_s^2\right>},\label{eq20}
\end{equation}
where $\left<G_{r\phi}\right>$ is the average vertically integrated $r$-$\phi$ component of the
gravitational stress tensor, $\left<R_{r\phi}\right>$ is the vertically integrated Reynolds stress tensor, $\Sigma$ is the surface density, $c_s$ is the midplane sound speed, and the brackets indicate that the quantities are azimuthally averaged \citep[see][]{gammie2001,lodato_rice2004}.  We may use the torque $C_z$ to find the stress from equations (16), (18) and (19),
\begin{equation} C_z=2\pi r^2\left(\left<G_{r\phi}\right> + \left<R_{r\phi}\right>\right).\label{eq21}\end{equation}
Using equations (18)-(21), we plot the time-averaged effective $\alpha$ for the 10-16 ORP time period in Figure 13. On the same figure, we show the $\alpha$ that is expected for a disk when entropy generation by gravito- and hydrodynamic turbulence provides enough heating to exactly balance cooling, namely, the $\alpha$ given by equation (2).  For the stretch between about 20 and 35 AU, Figure 8 shows that the $t_{\rm cool}=25/\Omega$ curve roughly overlaps  the average $t_{\rm cool}$ for the 10-16 ORP time interval.  If we take  $t_{\rm cool}\Omega\approx 25$ over that interval and calculate the $\alpha$  expected from equation (2), we find $\log \alpha = -1.9$ for the strongly self-gravitating case and $\log \alpha = -1.6$ for the negligible self-gravity case (see text below and equation [2]).  We also calculate an $\alpha$ based on the bold $t_{\rm cool}$ curve in Figure 8 with equation (2), which is averaged over 10-16 ORPs, and by assuming the negligible self-gravity case. In Figure 13, the $\alpha$ expected from equation (2) with the $t_{\rm cool}$ curve (heavy dashed curve) is consistent with the $\alpha$ derived from the gravitational torque with equation (20) (heavy solid curve) between 15 and 60 AU in $r$. This $\alpha$ is of the same magnitude as the $\alpha$ used in SED fitting for real disks \citep[e.g.,][]{hartmann1998}, but only in the 20-35 AU region is the effective $\alpha$, which is of the order $10^{-2}$, roughly constant.    

\subsubsection{Are the Torque Profiles Consistent with the Reported $\dot{M}$'s? }

To check the consistency of the reported $\dot {M}$'s with the torques in Figure 10, we calculate the torque that is required to produce the $\dot{M}$'s shown in Figure 3 for the 10-16 ORP interval.  With azimuthally averaged quantities so that each quantity only depends on $r$, 
\begin{equation} \dot{M} = \frac{2}{r\Omega}\frac{ {\rm d}}{ {\rm d}r }\left(C_z\right).\label{eq22}\end{equation}
Here, we have assumed that ${\rm d}\ln \Omega /{\rm d} \ln r=-3/2$, which is a very good approximation for most of our disk.  Equation (22) simply reflects the drift of mass due to changes in its orbital radius caused by torques. By integrating this equation, we calculate the total torque required to produce the $\dot{M}$'s measured in the RC disk.

We can view the consistency check from a different vantage point by using the torque from equation (22) to derive a corresponding $\alpha$.  We relate the torque and $\alpha$ by letting $\nu=\alpha c_s h$, by assuming that the disk scale height $h\approx c_s/ \Omega$, where the midplane values are used for $c_s$ and $\Omega$, and by using equations (20) and (21) in equation (22), which yields
\begin{equation} \dot{M} = \frac{6\pi}{r\Omega}\frac{ {\rm d}}{ {\rm d}r }\left(r^2\nu\Omega\Sigma\right),\label{eq23}\end{equation}

  The resulting torque profile from equation (22) and the $\alpha$ profile from equation (23) are shown as red curves in Figures 10 and 13, respectively.  This $\alpha$ profile is reasonably consistent with the $\alpha$ profile derived from the gravitational torque alone for the 5 to 50 AU internal.  Over that interval, the $\dot{M}$-derived $\alpha$'s are slightly larger than the gravitational torque-derived $\alpha$'s; this is probably due to ignoring constants of order unity in some of the relations. For $r>50$ AU, the $\alpha$'s deviate strongly.  
  
  In Paper II, we reported that the effective $\alpha$'s seen in our disks were an order of magnitude higher than those given by equation (2).  This is probably due to use of the mass flux in a steady-state, namely
  $\dot{M}=3\pi\nu\Sigma$, to compute the effective $\alpha$'s from the $\dot{M}$'s.  We suspect that the anomalously high effective $\alpha$'s in Paper II are incorrect for this reason; Michael et al. (in preparation) will examine the effective $\alpha$'s in constant $t_{\rm cool}$ disks with equations (20) and (23).   

\subsubsection{Summary}

The $\dot{M}$'s in the RC disk are primarily produced by global modes, as shown by Figures 10, 11, and 12, but the $\alpha$ profile derived from the torques and mass transport is consistent with a locally applicable $\alpha(r,t)$.  This is likely due to the close placement  of corotation radii for three strong $m=2$ modes over the 20-35 AU region.  As predicted by \citet{balbus_papaloizou1999}, a disk that is dominated by global modes will behave like an $\alpha$-disk as regards angular momentum transport and energy dissipation near the corotation radii of low-order modes.  We note that we are unable at this time to determine whether the energy dissipation in RC is consistent with some locally defined $\alpha(r,t)$, but we do find that the angular momentum transport is consistent.  We are currently addressing this issue, and plan to discuss it in a future paper.  Moreover, the disk deviates from an $\alpha$-disk behavior for $r> 50$ AU, which is far from the main $m=2$ corotation radii; this is consistent with \citet{balbus_papaloizou1999}.

Based on the the torque profile, the $\alpha$ profile, the $t_{\rm cool} \Omega$ profile, and the Fourier amplitude spectrum of the RC disk, we conclude that a local description for mass transport is consistent with an $\alpha$-disk near the corotation radii of low-order modes.  We stress, however,  that the {\it evolution} is not described by a constant $\alpha$.  Using an $\alpha(r,t)$ prescription is only possible if one knows the correct $t_{\rm cool}$'s for the asymptotic phase, which requires hydrodynamics calculations \citep[see][]{johnson_gammie2003}, and if one can estimate the location of corotation radii for low-order modes {\it a priori}.  In any case, the low-order modes lead to significant radial mass concentrations, e.g., Figure 4, which would be lost in a smooth $\alpha$ prescription.

\begin{figure}[ht!]
\begin{center}
\includegraphics[width=12cm]{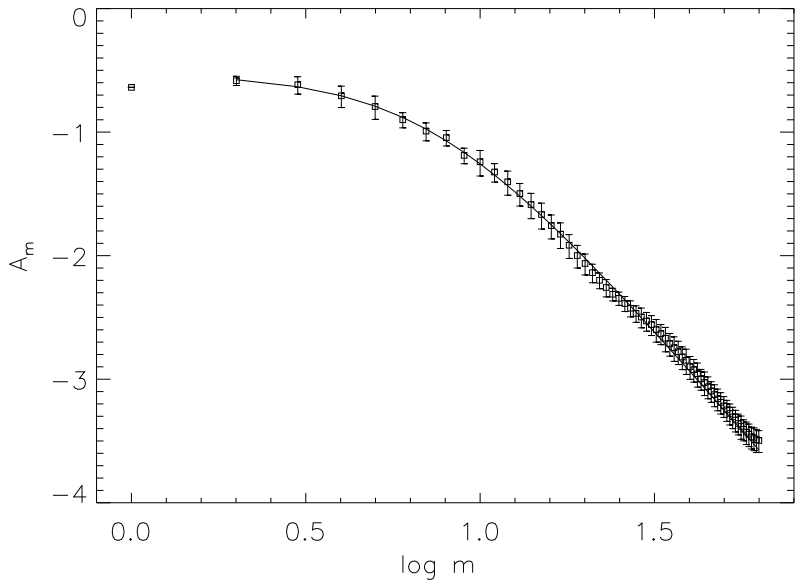}
\caption{The $A_m$'s vs.\ $m$ in log-log space.  The squares represent the data derived from the RC disk and the heavy line is the best fit curve through those data, excluding $m=1$.  The error bars represent the rms of the residuals and indicate the degree of fluctuation in the spectrum.  There is a noticeable wiggle in the linear portion of the curve, so the GI amplitude spectrum is probably more complex than described here. This is a topic for future investigation. }
\label{f12}
\end{center}
\end{figure}

\begin{figure}[ht!]
\begin{center}
\includegraphics[width=12cm]{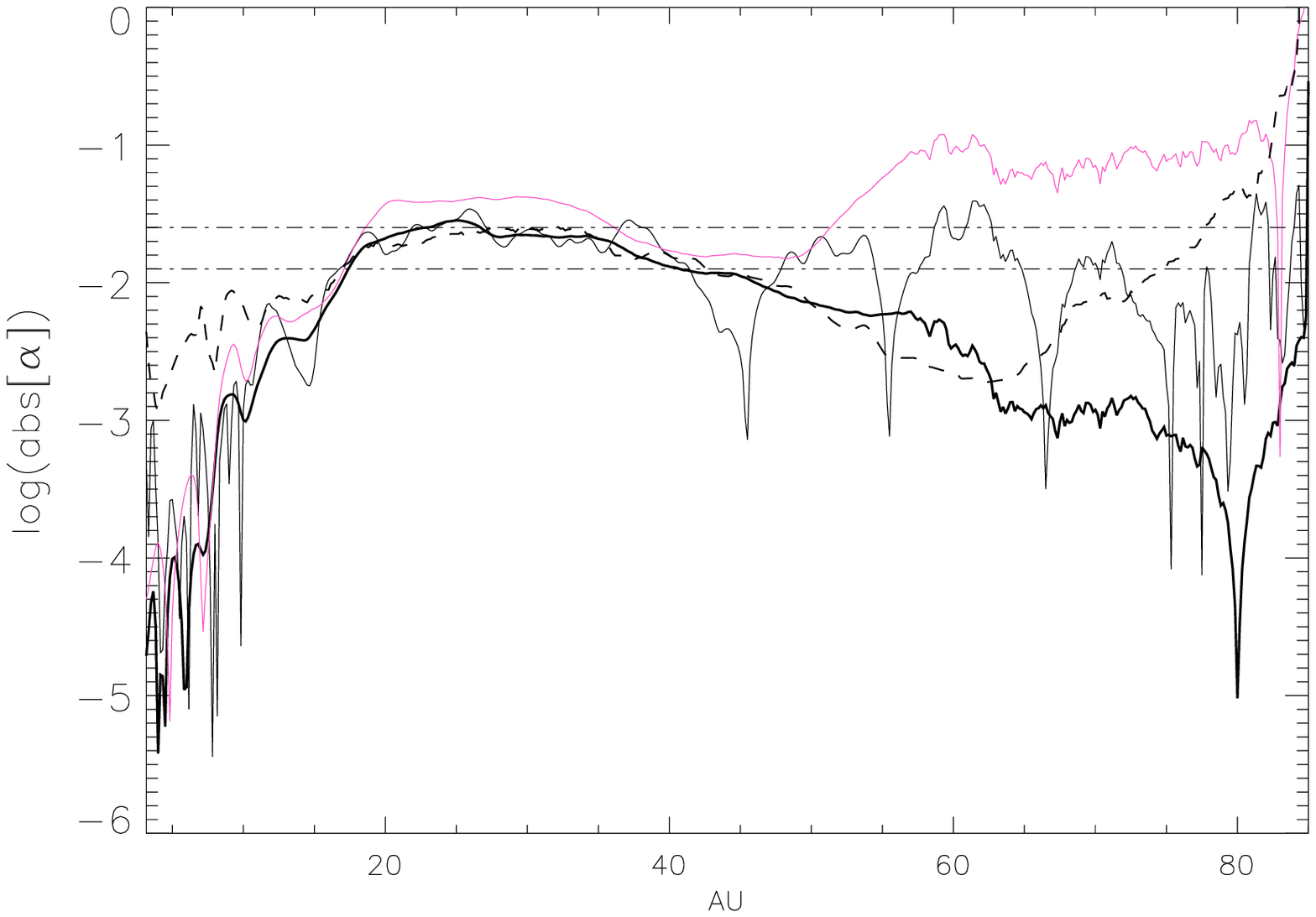}
\caption{Effective $\alpha$ based on the total torque profile in Figure 10 (thin solid curve) and the gravitational torque only (heavy solid curve).  The red [gray] curve indicates the $\alpha$ profile one would need to yield the $\dot{M}$'s for the 10-16 ORP interval shown in Figure 3. The dash-dot lines are effective $\alpha$'s predicted by Gammie (2001) for $t_{\rm cool} \Omega = 25$ for the strongly self-gravitating case (lower) and the nonself-gravitating case (upper).  The heavy dashed curve indicates the predicted $\alpha$ profile based on the $t_{\rm cool}$'s from the 10-16 ORP interval, which are shown in Figure 8, with the assumption of negligible self-gravity. }
\label{f13}
\end{center}
\end{figure}

\subsection{Fragmentation and Rings}

The RC disk develops a series of rings as massive as the ones seen in the CCT disk.  Matter is gathered at these radii and the rings are fed continuously, as seen in Figure 4.   This, along with the discussion in \S 5.1, implies that ring formation may be a robust result as long as there is a region of the disk with negligible torque or a region with kinks in the torque profile.

Recently, while conducting tests of this radiative scheme against a new scheme that implements rays for part of the radiative solution, it was noticed that the portion of the disk inside the first ring may be staying hot for numerical reasons.  This does not affect points made here about the rings, i.e., rings can build up where the torques precipitously decline.  In this case, the torques probably decline because, inside 7 AU, the disk is stable to GIs due to numerical heating.  However, there are several physical heating mechanisms that are excluded from this calculation, and we expect to see the same results if the inner 7 AU of this disk were to be kept hot for a physical reason.  We are currently running a series of calculations that do not have such a large radial range so that we may avoid poor numerics in the inner portion of the disk.

  \citet{durisen2005} speculate that concentrations of mass may contribute to planet formation by accelerating core accretion \citep[see also][Paper II \S 4.3]{pickett_lim2004}.  This might be the only way that GIs in RC could produce planets.  Typical cooling times in RC increase from fractions of an ORP to a few and even tens of ORPs with time (see Fig.\ 8).  Initially, $t_{\rm cool} \Omega \lesssim 6$ for several radii, but quickly (within 1 ORP) the cooling times increase, and the disk becomes strongly stable against fragmentation at all radii, because, as shown in Figure 8, $t_{\rm cool}\Omega  \gg 6$ \citep{gammie2001,johnson_gammie2003,mnras364l56}. This is also consistent with the stress fragmentation criterion \citep{mnras364l56}, which requires that the effective $\alpha$ in nonfragmenting disks be $\lesssim 0.06$.   The Toomre $Q$ drops below 1.0 only just before the burst phase at about $r=30$ AU.  To ensure that no fragmentation was overlooked due to resolution effects, the burst phase was rerun with 4 times as many cells in the azimuthal direction (512 instead of 128).  No significant structural change was observed.   Dense clumps do form during the burst phase at the intersections of spiral structures, but they last only a fraction of an ORP as in Papers I and II.  In addition to the burst phase, the asymptotic phase was extended for one ORP with the same high resolution. Qualitatively, the disk structure remains the same and no signs of fragmentation are detected.

The long cooling times in RC are consistent with the predictions of \citet{rafikov2005}, with cooling times in the \citet{nelson_benz_ruzmaikina2000} simulations \citep[see \S 4.2][]{durisen_ppv_chapter}, and with the cooling times due to radiative cooling alone in the \citet{boss2001,boss2002,boss2005} simulations.  However, Boss also claims to see convection in his simulations, which decreases the cooling times in his disks enough to allow for disk fragmentation and the formation of multi-Jupiter mass clumps.  We see no signs of convection in RC during the GI-active phases.

\subsection{Convection}

\citet{lin_papaloizou1980} \citep[see also][]{ruden_pollack1991} showed that a vertically contracting but otherwise quiescent protoplanetary disk will be convectively unstable when the optical depths are large, in the Rosseland mean sense, and when 
$\beta>\left( 3\gamma-4\right)/\left(\gamma - 1\right)$, where $\beta$ is defined as the exponent of the temperature dependence of mass absorption coefficient $\chi_{\rm Ross}$, i.e., $\chi_{\rm Ross} = \chi_{0}T^{\beta}$.  For a $\gamma=5/3$ gas, one should expect convection when $\beta>1.5$, and for $\gamma=7/5$, one should expect convection when $\beta>0.5$.   In our calculation, we find that $\beta\approx 2$ for $T\approx 30-100$ K, which is the temperature range in most of the interior (see Appendix A, Fig.\ 16).  Indeed, we see convection in the axisymmetric phase of the disk evolution during which the disk is, for the most part, undergoing quasistatic gravitational contraction in the vertical direction.  

We search for convection in the simulation by looking for convective ``cells'' that are associated with regions where the entropy gradient is negative, i.e., $\partial \left(P/\rho^{\gamma}\right)/\partial z < 0 $. We use the entropy gradient instead of the Schwarzschild criterion for convective instability, namely $\del = \partial\ln T/\partial \ln P > 0.4=\del_{\rm ad}$ for a $\gamma=5/3$ gas, because the Schwarzschild criterion assumes a vertically hydrostatic background, which is not guaranteed in a dynamic disk.  

Figure 14 shows superadiabatic regions in the disk at 1.2 ORPs (top panel) and at 10.6 ORPs (bottom panel).  Both panels show heating due to artificial viscosity by gray-filled contours, velocity vectors scaled to each axis but with the heads on the left panel being slightly larger for clarity, and heavy curves representing density contours.  The superadiabatic regions in the right panel are delineated by blue thin curves. The left panel omits these contours to show the velocity vectors more clearly; almost the entire region inside the inner most density contour is superadiabatic.  In the left panel, the average Mach speed in the convective eddies $\left< v_z/c_s \right> \approx 0.06$, where $\left< v_z/c_s \right> 
= \int  \rho~\left| v_z\right| /c_s~{\rm d}V/\int \rho~{\rm d}V$ and the integrals are evaluated over the volume spanning between 15 and 25 AU in $r$.  The value of $\left<v_z/c_s\right>$ fluctuates between 0.01 and 0.15 when each annulus of grid cells is evaluated separately ($\Delta r= 1/6$ AU), and  some of the convective eddies result in nontrivial compressional heating through artificial viscosity.   To evaluate energy transport by convection, we estimate the convective flux ${\rm F}_c = -1/2 c_p \rho T v_z \ell \left({\rm d}\ln T/{\rm dz} -\del_{\rm ad} {\rm d}\ln P/{\rm d}z\right)$ \citep{lin_papaloizou1980} through each cell in the volume described above. Here, the mixing length $\ell={\rm min}\left(z,P/\Omega_k^2 z\rho\right)$ and $c_p$ is the specific heat at constant pressure. By dividing the total internal energy within the half disk between 15 and 25 AU in $r$ by the total convective energy loss rate for that same region, we find that the convective cooling time is about 1 ORP and that it is comparable to the radiative cooling time.  However, this method measures the convective flux based on a superadiabaticity estimate according to the formalism of \citet{lin_papaloizou1980}. For a second measurement, we calculate the energy carried by convective motions through a plane that cuts through the volume at $z\approx1$ AU. The convective flux through the $i$th cell is ${\rm F}_{c}|_i = \rho v_z c_p \Delta T |_i$, where $\Delta T$ is the difference between the actual temperature at the cell center and the azimuthally averaged temperature for that $r$ and $z$.  We find a convective cooling time of about 2 ORPs with this method.  According to either estimate  of the convective flux, both of which are crude and uncertain, convection is efficient at redistributing energy in the disk's interior. However, these high convective fluxes may be due to a combination of the random perturbation to the initial constant vertical entropy profile, of our seeding superadiabatic regions at the interior/atmosphere interface (see Appendix B), and of our exclusion of irradiation, which tends to produce stabilizing temperature stratification.  Moeover, this energy ultimately must be {\it radiated} away, so the cooling times in Figure 8, based on energy actually radiated, accurately reflect disk cooling times.   

The convective cells that are established in the axisymmetric phase are disrupted during the burst phase, and, for the subsequent GI-active phases, we do not see convection.  We believe this is due to two effects: (1) The dynamics in the disk, namely spiral waves, disrupt any establishment of convective cells due to the large fluctuations in mass transport.  (2) These spiral shocks lead to shock bores, which in turn lead to high temperatures at 
moderate and high disk altitudes \citep[see also Fig.\ 7, here]{boleyshockbores}.  These high-altitude shocks should make the disk convectively stable by keeping the entropy gradient of the disk positive or essentially zero in the vertical direction.  The right panel in Figure 14 shows this well.  Despite the presence of superadiabatic gradients (see appendix B for an explanation why there are strong superadiabatic regions at mid-disk altitudes), convective cells are absent.  The only vertical motions associated with the flow are due to shock bores along the spiral shocks or other waves.  

Because shock bores deposit energy into the upper layers of the disk, it is reasonable to ask whether shock bores themselves provide convective-like cooling.  Although this point is still unclear, we believe not.  Shock bores effectively reduce the temperature in the post-shock region as the jumping gas expands vertically.  This removes thermal energy in the post-shock region and deposits that energy in the upper layers via waves.   Therefore, we do expect shock bores and possibly other wave effects \citep[e.g., see][]{lubow_pringle1993,lubow_ogilvie1998,ogilvie1998} to enhance disk cooling when large-scale spiral shocks are present in the disk.  Although energy may be effectively removed from the post-shock region, the interior of the disk can still only cool by radiative diffusion because of the positive vertical entropy gradient established by the shock bores. From this point of view, shock bores and other wave effects limit the efficacy of spiral waves in heating the disk.  Distinguishing between shock bores and convection is not a point of semantics.  Shock bores are born of large-scale shocks in a disk, while thermal convection and the criterion for convective instability are usually described in the context of a disk in vertical hydrostatic equilibrium. 

\begin{figure}[ht!]
\begin{center}
\includegraphics[width=15cm]{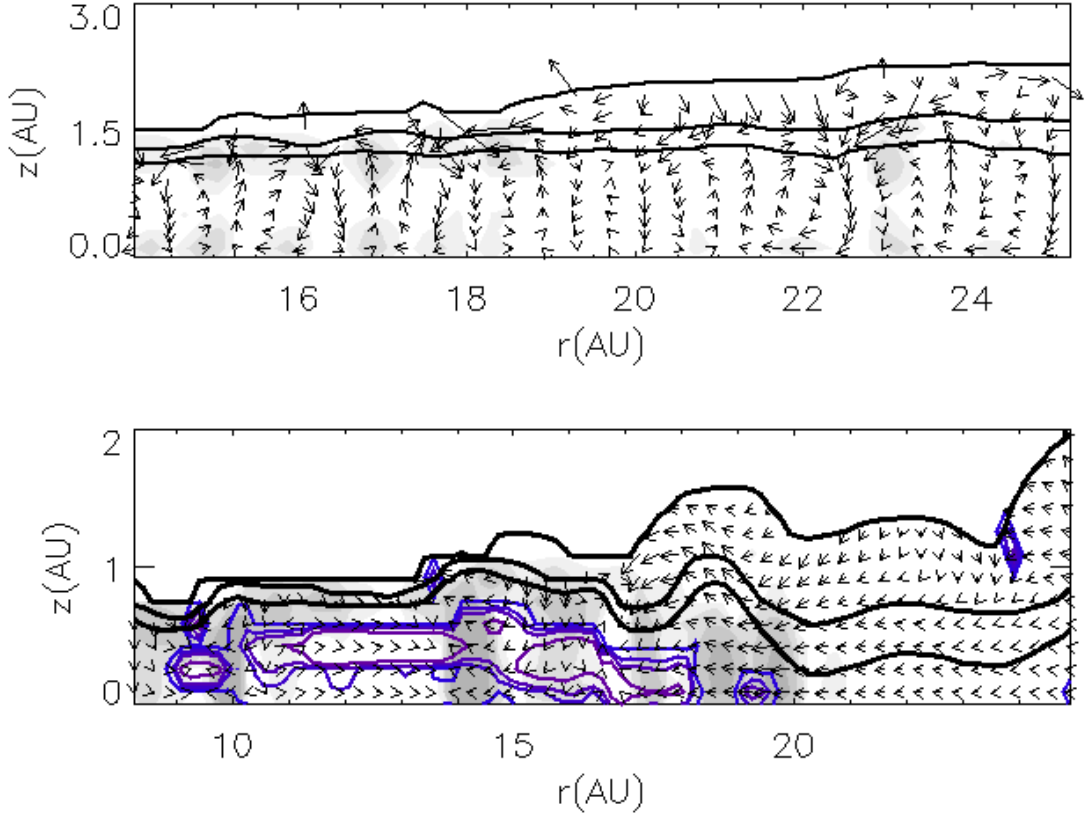}
\caption{The dark lines indicate density contours, the gray shading indicates heating from bulk viscosity, and the thin blue/purple lines indicate superadiabatic regions, which are defined wherever the vertical entropy gradient is negative.  The arrows show the direction of gas flow in this azimuthal slice.  Top: An $r$-$z$ slice through the disk during the axisymmetric phase (1.2 ORPs) when convection occurs.  The entire region below the innermost density contour is superadiabatic, so the contours are omitted.  Bottom: A slice through the disk but at 10.6 ORPs and for a larger region.    Convection is absent in the superadiabatic regions; gas that is rapidly moving upward is related to shocks in the disk and are partly shock bores.  The velocity vectors have a different scaling because the velocities are large.}
\label{f14}
\end{center}
\end{figure}

\subsection{Spectral Energy Distributions}

As emphasized by \citet{nelson_benz_ruzmaikina2000}, an important test of physical relevance is a comparison between SEDs of observed systems and the derived SED from a simulation.  We attempt to construct what the RC disk would look like to an observer at some large distance $d$ looking at the disk face-on.  For simplicity, we assume that each $(r,\phi)$ column of the disk emits radiation according to a black body law at its effective temperature over its surface area.  Because we use a cylindrical grid, the $i$th area element has a solid angle ${\rm d}\Omega_i=r_i{\rm d}r{\rm d}\phi/d^2$.  The specific flux, or flux density, then can be tallied by
\begin{equation} {\rm F}_{\nu}=\sum_i {\rm d}\Omega_i B_{\nu}\left(T_i\right),\label{eq24}\end{equation}
where $B_{\nu}$ is the Planck function and $T_i$ is the effective temperature of the $i$th cell. To avoid using distance, we choose to express the SED in terms of $4\pi d^2\nu {\rm F}_{\nu}$.  As a basis for comparison, we adopt the approach of \citet{nelson_benz_ruzmaikina2000} and define our fiducial to be an SED derived from a disk with a temperature law $r^{-0.6}$, which is the median best fit law to the T Tauri disk sample presented in \citet{beckwith1990}.  Figure 15 shows the SEDs derived from RC for three different stages in its evolution.  The short dashed line delineates the SED for the disk near its brightest period during the burst phase (2.4 ORPs), when the luminosity is nearly 19 $L_{\odot}$ as integrated between $10^{10}$ and $10^{15}$ GHz.  The long dashed line delineates the SED for the disk as it enters the asymptotic phase (10 ORPs), and the solid black line indicates the SED for the disk at the end of the calculation (16 ORPs).  The star, which is assumed to have an $R=2R_{\odot}$ and a $T_{\rm eff} = 4,000$ K, is included in the SED profile; the SED has dips in specific luminosity as it nears the star because it has a 2.3 AU hole and is missing a contribution from an inner and hotter portion of the disk.  The red lines indicate fiducial SEDs based on assuming a $T_{\rm eff}\sim r^{-0.6}$ temperature profile for a 0.0033, 0.01, and 0.033 $L_{\odot}$ disk and integrating between 0 and 60 AU.  The fiducial SEDs also have a slight dip in their profile just before transitioning to the stellar portion of the SED due to discretizing the temperature profile into grid cells 1/6 AU wide in $r$.  The blue line is a fiducial SED calculated for a luminosity of $0.0024~L_{\odot}$, which is the disk luminosity at 16 ORPs, in the same way as the other fiducials but with a 2.3 AU hole.  Even though the actual effective temperature profile for RC is an exponential (see Fig.\ 6), the SED for RC is observationally consistent with a profile $T_{\rm eff}\sim r^{-0.6}$ because the inner 20 AU of the disk closely follows a $T_{\rm eff}\sim r^{-0.59}$.   Although GIs appear to lead to an exponential $T_{\rm eff}$ profile for this disk, stellar irradiation \citep{dalessio2001} probably keeps the outer regions of the disk warmer than what is modeled here inasmuch as that region of the disk is flared (see Fig.\ 9).

\begin{figure}[ht!]
\begin{center}
\includegraphics[width=15cm]{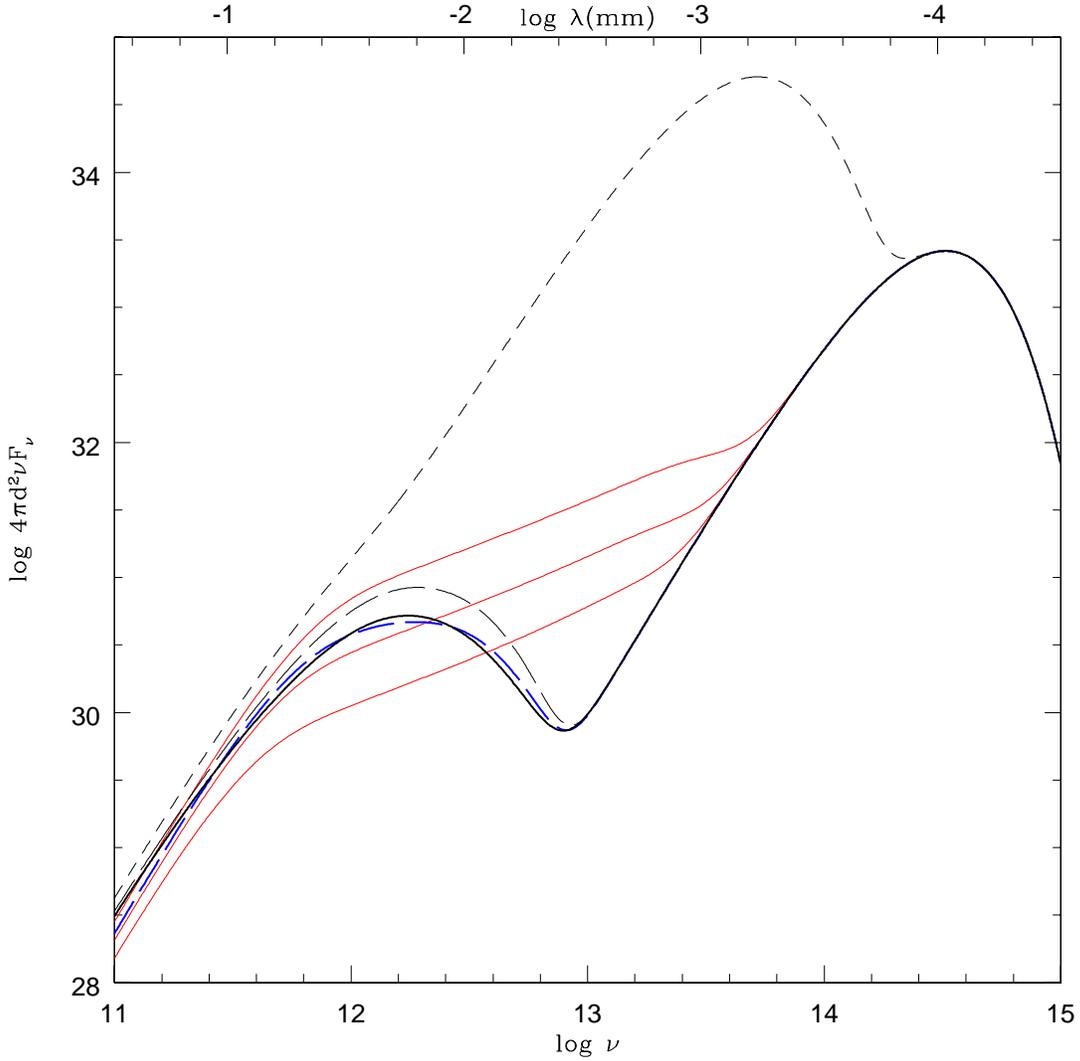}
\caption{The solid line indicates the SED for RC at the end of the calculation (16 ORPs).  The short dashed and long dashed curves delineate the SEDs for RC at 2.5 ORPs and at 10 ORPs.  The red [gray solid] curves represent fiducial SEDs for 0.0033, 0.01, 0.033 $L_{\odot}$ for a disk that extends from 1/6 AU to 60 AU and has a $T_{\rm eff}\sim r^{-q}$ for $q=0.6$.  The blue [dark] dashed curve indicates the SED for a disk that is truncated at 2.3 AU with a $q=0.6$.  }
\label{f15}
\end{center}
\end{figure}

The SEDs of RC also suggest that an FU Ori outburst may be the signature of a disk becoming gravitationally unstable. The RC disk reaches a peak luminosity of 19 $L_{\odot}$ at about 2.4 ORPs, which makes RC approximately 10,000 times brighter in the burst phase than in the asymptotic phase.  Furthermore, the disk has a luminosity of only about 0.01 $L_{\odot}$ around 2.1 ORPs, so the disk luminosity increases by a factor of about 1,000 in 80 years.  The increase in luminosity is sudden and may be much shorter than we are reporting because our time resolution for determining disk luminosity is only about 0.3 ORPs due to data handling restrictions, which also means we may be missing the peak brightness.  GIs occur on a dynamical time scale and the steeping of spiral waves into shocks may be rapid enough to be consistent with observed FU Ori outbursts, which have luminosity rise times between a few to tens of years \citep{hartmann1996}.  In addition, the decay time scale for RC is about one ORP (about 250 yr), which is consistent with observed decay times.  

Although the burst phase is a promising explanation for FU Ori outbursts, one may certainly ask the question: Is the burst phase an artifact of our initial conditions?  The answer is unclear.  The burst phase is the transition from a globally stable disk to a globally unstable disk.  We speculate that the burst we see in this disk might be applicable to a  real disk and to FU Ori outbursts under the following scenario: A real disk might form a dead zone \citep{gammie1996,armitage2001} wherein GIs may undergo episodic bursts. These bursts of GIs in an annulus can create strong shocks over a large $\Delta r$ \citep{boleyppv} and produce FU Ori outbursts with very short rise times.  The location of the unstable annulus determines the rise time for the outburst.  Once the disk becomes unstable, the shocks extend deep into the inner disk, which is excluded from this calculation and should lead to even brighter bursts with more luminosity at shorter wavelengths. The global torques exerted on the disk as a result of the burst are about 10 times greater than in the asymptotic phase, which results in effective $\alpha$'s that exceed $10^{-1}$ without leading to long-term fragmentation.  The strong shocks and redistribution of mass lead to the restabilization of the unstable annulus and of the disk as a whole, and the increase in the envelope irradiation will also work to stabilize the disk \citep{cai_letter_2006}.   Furthermore, an annulus of material with a mass around 10 Jupiter masses is expected to be completely redistributed in the nebula in a few hundred to a few thousand years, assuming a mass flux of between  $10^{-5}$ and $10^{-4}$ M$_{\odot}$ yr$^{-1}$. More detailed modeling is necessary, but we believe that there is enough evidence to suggest that FU Ori outbursts may be the signature of the sudden onset of GIs in protoplanetary disks.  If this is true, then FU Ori outbursts may also signal the annealing of dust and  the formation of chondrules in protoplanetary disks by spiral shocks \citep{wood1996,bossdick,boley2005proc,boleyppv}.

\subsection{Fate of the Disk}

The internal energy profile in Figure 5 shows that, after the burst, the disk transitions to a self-sustaining marginally unstable disk, but, as discussed in \S 3.4, that phase can only be maintained for a few $\times 10^4$ yr.  Because this is an incomplete model, i.e., we exclude the inner disk, irradiation, possible infall, dust settling, grain growth, and other mass transport mechanisms, it is difficult to comment on the disk's ultimate fate.  Nevertheless, we speculate about the bigger picture, and hope that it might serve to link several areas of disk research.

The time scale for grain growth from a maximum size of about 1 $\mu$m to about 1 mm is expected to be between the orders $10^3$-$10^5$ yr \citep{haghighipour2005}.  This time scale is commensurate with the lifetime of the asymptotic phase of the RC disk, {\it ceteris paribus}, and it therefore must have an effect on the disk's evolution.  Even though our disk has a relatively high effective $\alpha$, it is unclear whether there is enough turbulence in the disk to keep the larger mm-size grains from settling to the midplane. To proceed, we assume that no significant settling takes place while the GIs are as active as reported in \S 5.  Under this scenario, the larger grain size will increase the opacity in cooler portions of the disk where the typical photon wavelength is long, while the inner disk will see a decrease in opacity where the typical photon wavelength becomes shorter than the maximum grain size \citep[see, e.g.,][]{dalessio2001}.  We expect that, in the cool parts of the disk, the cooling times will increase as the maximum grain size grows; this is consistent with the simulations presented by \citet{cai_letter_2006}.  Longer cooling times lead to weaker GIs and to slower evolution.  As grains grow, the GIs may become so weak that they no longer provide enough turbulence to counter settling.  Because the dust settling time is similar to the grain growth time \citep{weidenschilling1997,haghighipour2005}, dust settling is rapid.  As the dust settles to the midplane, the opacity for most of the disk decreases, which would decrease the cooling times.  This would lead the gas disk toward stronger instability.  If the instability is strong enough, the disk may undergo multiple burst-like phases, during which the effective $\alpha$'s $\sim 10^{-1}$.  Because rings are a pressure maximum that lead to enhanced dust to gas ratios \citep{weidenschilling1977sn,weidenschilling1997,haghighipour_boss2003,rice2004,durisen2005,haghighipour2005}, if rings form at the outer radius of a dead zone, one might expect grain growth and dust settling to happen there first.  Whether the entire outer disk becomes unstable or just a large annulus, these bursts could re-mix the grains vertically and radially in the disk \citep{boley2005proc,boleyshockbores} and lead to collisional grain destruction and to the processing/reprocessing of crystalline material.  Because bursts redistribute the mass in the disk, the lifetime of the disk increases.  

An opacity increase from grain growth can lead to a cessation of GI activity. The cessation of strong gravitoturbulence permits dust settling if there are no other means for producing strong turbulence.  This dust settling might decrease the opacity enough that the cooling times become short enough to induce another disk burst. This sequence may be able to explain FU Ori outbursts and their duty cycle, drive shocks for chondrule formation \citep[e.g.,][for a discussion on chondrule formation via shocks]{desch2002}, and might extend the RC disk lifetime to $\sim 10^6$ yr.  Although speculative, we believe this provides an avenue for further study.

 \section{CONCLUSIONS}
 
Our simulations of a protoplanetary disk show that (1) radiative cooling times are too long to permit disk fragmentation for a $\gamma = 5/3$ when realistic opacities are used (2) mass and angular momentum transport are dominated by low-order modes, (3) angular momentum transport in gravitationally unstable disks with realistic cooling is consistent with the transport expected for an $\alpha$-disk over a nontrivial region of the disk, (4) {\it evolving} a GI-active disk as if it were an $\alpha$-disk may be impossible to implement correctly because it requires knowing the $t_{\rm cool}$'s and the distribution of corotation radii for the low-order modes {\it a priori}, (5) the GI activity in our disk is weaker than in the constant cooling time disks of \citet{mejia2005} because the cooling times are typically longer, (6) the duration of the GI-active phase in this disk is expected to continue for a few $\times 10^4$ yr, (7) convection is present during the axisymmetric phase but convective cells are completely disrupted by GI-activity, and (8) features in the torque distribution can lead to the development of rings near the edge of a GI-active region. We also find that our simulation yields an SED that is compatible with observed SEDs and that the GI burst phase may correspond to an FU Ori outburst-like event.  
 
The behavior of gravitational instabilities in disks is extremely sensitive to the handling of thermal physics, and researchers must use extreme caution when treating radiative boundary conditions.  We find most of the disk volume is in the atmosphere, which is optically thin to its own radiation, as also found in the \citet{cai_letter_2006} simulations.  Therefore, proper disk evolution depends on the treatment of radiative cooling at low optical depths.  Cooling times tend to increase past the initial phase to the point that a vertical column can take from several hundreds to tens of thousands of years to radiate its internal energy.  These time scales agree with those seen by Boss (2002) due to radiative cooling alone and are too long to produce planet formation by disk fragmentation.  Convection could provide shorter cooling times \citep{boss2002}, but convective motions are absent in the simulations presented in this paper and in \citet{cai_letter_2006} during GI-active phases.  Even if the shock bores that produce vertical motion in our disk enhance the cooling, the cooling times are still long.

Assumptions about the physical properties of dust grains that are mixed with the gas in these disks play an important role in determining the cooling rates of the entire disk.  The problems of dust settling \citep{calvet2005} or aggregation of solids \citep{rice2004} are not addressed in these models, but they will be significant factors to consider in future simulations.  Moreover, a $\gamma=7/5$ gas disk may behave differently from what is modeled here \citep{mnras364l56}.  Including all the details necessary to handle the flow of energy in circumstellar disks not only requires a large amount of future human effort and computing time, but it is also requires better constraints on the physical parameters of real disks.  We are now developing a new radiative scheme that couples ray tracing with radiative diffusion in an effort to improve coupling of optically thick and thin regions.

\acknowledgments{We thank P.~Armitage, A.~Boss, N.~Calvet, C.~Gammie, T.~Hartquist, G.~Lodato, S.~Lubow, S.~Michael, \AA.\ Nordlund, J.~Pringle, and W.~Rice for useful comments, suggestions, and conversations related to this research. We also thank the anonymous referee whose comments helped improve this manuscript.  A.C.B.~was supported by a NASA Graduate Student Research Program fellowship. A.C.M.~acknowledges the NASA Astrobiology Institute for support.
R.H.D.'s and K.C.'s contributions were supported by 
NASA grants NAG5-11964 and NNG05GN11G.  M.K.P.~was supported by a grant from NASA's Planetary Geology \& Geophysics Program and the visitor program at the University of Leeds.}  

\appendix

\section{OPACITIES}

All the opacities used in the radiative cooling simulations were obtained from Paola D'Alessio.  The opacities are similar to those by Pollack et al.\ (1994), but with a few modifications (see D'Alessio et al. 2001 for details).  The main contributors are grains of H$_2$O ice, silicate (olivine [Fe,Mg]$_2$SiO$_4$ and orthopyroxene [Fe,Mg]SiO$_3$), organic (containing C, H, O, and N), and troilite (FeS) (Figure 16).  The dust grains are assumed to be spherical, and their size distribution is a power law
\begin{equation} dn = n_{0}a^{-s}da,\label{eqA1}\end{equation}

where $a$ is grain radius, $n_{0}$ a normalization constant, and $s$ a free parameter.  The opacities used in all the simulations assume $s = 3.5$, which best fits observed ISM extinctions for grains of various compositions \citep{mathis1977,pollack1994}.  The minimum and maximum grain radii are 0.005 and 1 $\mu$m, respectively, although opacity tables for different maximum grain radii, $a_{\rm max}$, are available.  While these grain sizes are more representative of interstellar dust \citep{draine_lee1984}, they have been widely used to model composition, abundances, and physical properties of the Solar Nebula and circumstellar disks in general \citep[e.g.,][]{pollack1985,pollack1994,calvet1991,henning_stognienko1996,bell1999,dalessio1999}.  This appendix gives the definitions of the various mean opacities used in the 3-D hydrodynamics code, summarizes their properties, and briefly compares how different grain sizes can affect disk structure and energy transport.  For details on the specific opacities that are used to calculated the mean opacities described in this appendix, we refer the reader to \citet{dalessio2001}.

Since the 3-D hydrodynamics code does not compute frequencies or wavelengths, the opacities used in the calculations of the heating and cooling terms in (\S 2.1) are the Planck and the Rosseland mean opacities integrated over all frequencies, defined as
\begin{equation} \kappa_{\rm Planck} = \frac{ \int_0^{\infty} \kappa_{\nu} B_{\nu} {\rm d}\nu}
{ \int_0^{\infty} B_{\nu} {\rm d}\nu},\label{eqA2}\end{equation}
and
\begin{equation} \chi_{\rm Ross} = \frac{ \int_0^{\infty} \frac{\partial B_{\nu} }{\partial T} {\rm d}\nu }
{  \int_0^{\infty} \frac{1}{\chi_{\nu}} \frac{\partial B_{\nu} }{\partial T} {\rm d}\nu },\label{eqA3}\end{equation}
respectively, where $\kappa_{\nu}$ represents only true absorption while $\chi_{\nu}$ constitutes total extinction, i.e., absorption plus scattering and where $B_{\nu}$ is the Planck function.   Rosseland opacities are used in optically thick regions, where the diffusion approximation is valid and the mean free path of a photon is much smaller than the thickness of the gas.  Planck opacities better characterize regions where photons are likely to escape without interacting with the medium \citep[e.g.,][]{mihalas_mihalas}.  Figure 16 shows $\chi_{\rm Ross}$ and $\kappa_{\rm Planck}$ vs.\ temperature for maximum grain sizes $1\mu$m and 1mm.

\begin{figure}[ht!]
\begin{center}
\includegraphics[width=12cm]{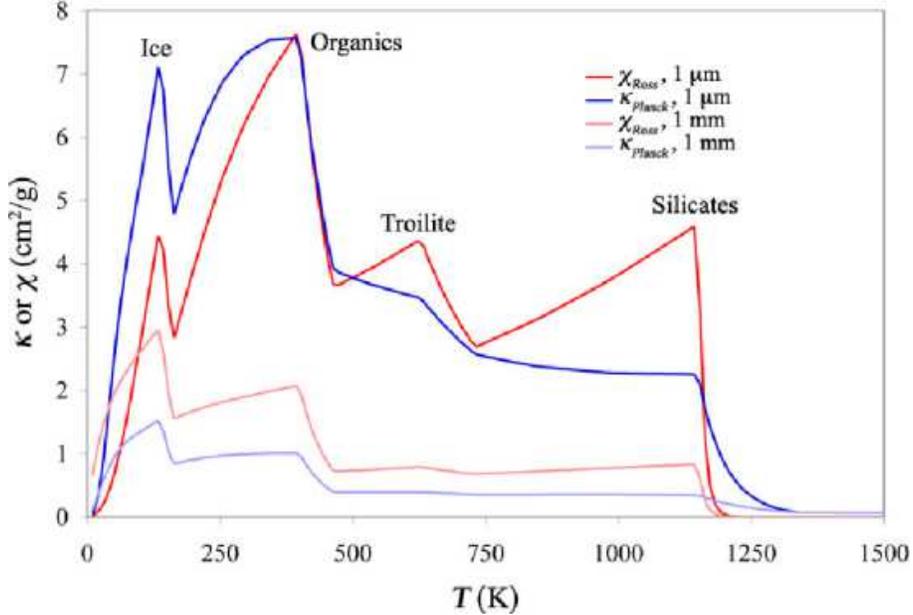}
\caption{Planck and Rosseland mean opacities vs.\ temperature for $a_{\rm max} = 1 \mu$m (solid heavy curves) and 1 mm (lighter curves).}
\label{f16}
\end{center}
\end{figure}

The dips seen in Figure 16 are due to the vaporization temperatures of the different dust constituents.  For a typical number density of the initial model, about $3\times10^{11} \rm ~cm^{-3}$, water ice is the main contributor to the opacities under 160 K.  At higher temperatures the ice is vaporized, so most of the extinction is due to organic grains ($< 470$ K), troilite ($< 740$ K), and silicates ($< 1140$ K).  Notice that for $a_{\rm max} = 1 \mu$m the Planck mean opacities are larger than the Rosseland mean opacities in the temperature regime of the simulations ($T < 300$ K).  This means that for the RC disk, the first term on the right hand side of equation (7) is likely to be larger than the second term, so the net cooling of the atmosphere is positive.  On the other hand, the Rosseland opacities are larger than the Planck opacities for $a_{\rm max} = 1$ mm in the same temperature range.  In this case, the atmosphere is likely to have negative cooling or net heating in equation (7), which, when added to heating by shocks, can cause the atmosphere and the disk to expand very quickly and unphysically, as was observed in some preliminary tests.  To avoid this problem, the calculations presented in \citet{cai_letter_2006} use the Planck mean opacities for all terms in equation (7).

The opacities used here are a function of grain size. Smaller grains contribute more to the opacities at temperatures of several hundred Kelvin, while larger grains are the main contributors for temperatures lower than 100 K.  This indicates that the initial vertical physical thickness of the atmosphere in the disk models is determined by the choice of $a_{\rm max}$.  Tests show that the atmosphere is physically thinner for $a_{\rm max} = 1$ mm (higher column optical depths)  for the temperature ranges of the initial model (tens of Kelvin in the mid and outer disk).  This accounts for the longer overall cooling times and slower evolution of the $a_{\rm max} = 1$ mm simulation in \citet{cai_letter_2006}.  The energy transport processes in disk simulations depend significantly on the grain size adopted for the opacities.

\section{RADIATIVE TEST}

As argued here and in \citet{cai_letter_2006}, different treatments of boundary conditions could drastically change the behavior of the simulation.  Although an analytic solution for radiative hydrodynamics in a disk does not exist, there are several simple test cases and approximations \citep[e.g.,][]{hubeny1990} that can be explored as a means to test the accuracy of any particular radiative physics scheme.  We present one such test case in this appendix and challenge all researchers who publish radiative hydrodynamics simulations to perform similar tests or to develop tests of their own and publish the results.

The test we present here checks the code's accuracy in achieving a plane-parallel gray atmosphere solution.  Assume the opacity $\kappa$ is constant, so that
\begin{equation} \tau\left(z\right) = \int^{\infty}_z {\rm d}z'~\rho\left(z'\right)\kappa=m\left(z\right)\kappa,
\label{eqB1}\end{equation}
where $m(z)$ is the surface density down to $z$.  Furthermore, by adopting a constant gravitational accleration $g$, vertical hydrostatic equilibrium requires a pressure profile that follows
\begin{equation} P(z)=gm(z),~P(\tau)=g\frac{\tau}{\kappa}.\label{eqB2}\end{equation}
If we require that flux be conserved, then the temperature profile should be close to that of the standard Eddington approximation, i.e.,
\begin{equation} T(\tau)=T_{\rm eff}\left[\frac{3}{4}\left(\tau+q\right)\right]^{1/4},\label{eqB3}\end{equation}
where $q=2/3$ classically (Eddington approximation) but is closer to $q=1/\sqrt 3$ for real atmospheres \citep{mihalas_mihalas}.  For the test simulation, we place a slab of material, which is in hydrostatic, but not radiative, equilibrium vertically in our cylindrical grid with a low resolution of $(32, 8, 64)$.  We freeze the radial direction to avoid boundary condition problems at the ÒdiskÓ edges.  In this vein, the test calculation is two-dimensional hydrodynamically, but three-dimensional radiatively.  The radial flux at the edges of the disk is set to zero to conserve vertical flux.  A  flux $\sigma T_{\rm eff}^4$ is introduced in the vertical direction at the midplane, where $T_{\rm eff}$ = 800 K for the test.   Moreover, we set the $\kappa_{\rm Planck}=\chi_{\rm Ross}=\kappa=1\rm ~cm^2/g$ and $m(0) = 20\rm ~ g/cm^2$ so that $\tau_{\rm midplane}  = 20$.
 
	Figure 17 shows that the temperature structure deviates no more than about 6 \% from the analytic solution, with $q=2/3$, in the region where  $\tau>2/3$.  The boundary condition we impose allows for the correct flux, as calculated by the flux-limited diffusion routine, through the disk interior to within a percent despite the error in the temperature.  Unfortunately, the atmosphere does experience a sudden drop in temperature above the photosphere.  This drop is due to the lack of the complete cell-to-cell coupling in equation (7) that radiative transfer requires. As seen in Figure 17, we expect the RC disk's atmosphere to have a temperature drop that is comparable with the drop in this test or even larger due to the differences in the Rosseland and Planck means.  Because we calculate the boundary flux accurately, the temperature drop should not be a problem for disk cooling.  The largest problem that might be introduced by this sudden drop is dynamical.  Recall that the entropy  $S\sim\ln\left(P/\rho^{\gamma}\right)\sim\ln\left(T/\rho^{\gamma-1}\right)$, so a sudden temperature drop without a balancing density drop will result in a negative vertical entropy gradient.  As a result, our atmospheric fitting routine apparently seeds superadiabatic regions in our disk where the interior and atmosphere meet (Fig.\ 14).  It should be noted that, even so, we do not see thermal convection during the GI-active phase and that we do not expect these superadiabatic regions to result in thermal convection because the optical depths are small.
	
Our scheme, though crude, accomplishes everything we require:  Most importantly, we estimate the photospheric flux from the disk interior reasonably well.  Although the atmosphere, which is optically thin, is too cool, we do allow cooling in the atmosphere through the first term in equation (7), and we prevent the atmosphere from contracting completely onto the disk by adding the second term in equation (7).  It is necessary to include the atmosphere in the cooling treatment at some level of approximation because dynamics in this region create thermal effects \citep{boleyshockbores} and because we treat cases with external irradiation \citep{cai_letter_2006}.  We are developing additional tests to assess time-dependent thermal fluctuations.  

\begin{figure}[ht!]
\begin{center}
\includegraphics[width=12cm]{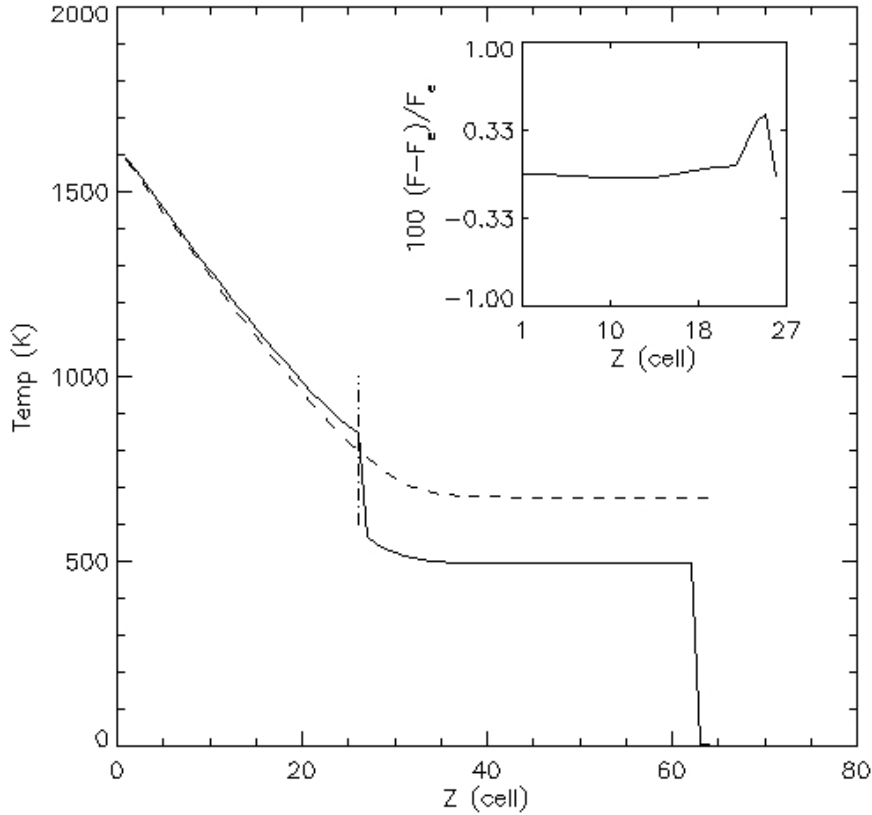}
\caption{Temperature profile for the test simulation.  The temperature structure for the interior disk (black curve) matches the analytic value (dash curve for $q=2/3$) fairly well.  The atmosphere goes to an isothermal profile, as expected, but it is too cold.  The vertical line shows the approximate location of $\tau=2/3$. The inset of this figure shows the percent change in the flux as calculated by the flux-limited diffusion routine from the Eddington flux $\rm F_e$ that is introduced at the base of the atmosphere. }
\label{f17}
\end{center}
\end{figure}

\clearpage

\bibliography{chonbib}
\bibliographystyle{apj}

 \end{document}